\documentclass[iop,preprint]{emulateapj}
\usepackage{ulem}
\usepackage{amsmath}
\usepackage{subfigure}

\def\INTEGRAL{\textit{INTEGRAL}}
\def\INTEGRALSPI{\textit{INTEGRAL} SPI}

\def \phf{ph\ cm$^{-2}$\ s$^{-1}$ }

\def \Ared4.9{A[${4.9\mu}$]}

\def\COMPTEL{\textit{COMPTEL}}
\def\RHESSI{\textit{RHESSI}}
\def\SPI{SPI}
\def\EGRET{\textit{EGRET}}
\def\HEAOC{\textit{HEAO-C}}
\def\dixmoinscinq{$\times$ 10$^{-5}$ ph\ cm$^{-2}$\ s$^{-1}$}

\def\Al{\(^{26}\)Al}
\def\Fe{\(^{60}\)Fe}

\def\Bouchet11{paper I}

\slugcomment{Draft version \today, Accepted for publication in The Astrophysical Journal}
\shorttitle{Galactic \Al\ emission with \INTEGRALSPI}
\shortauthors{Bouchet et al.}

\begin{document}
\title{The Galactic \Al\ emission map as revealed by \INTEGRALSPI}

\author{Laurent Bouchet\altaffilmark{1,2}, Elisabeth Jourdain\altaffilmark{1,2}, 
and Jean-Pierre Roques\altaffilmark{1,2}} 
\email{lbouchet@irap.omp.eu}
\altaffiltext{1}{Universit\'e de Toulouse, UPS-OMP, IRAP,  Toulouse, France}
\altaffiltext{2}{CNRS, IRAP, 9 Av. colonel Roche, BP 44346, F-31028 Toulouse 
cedex 4, France}

\begin {abstract}
Diffuse emission is often  challenging since it is undetectable by  most instruments, which are generally 
dedicated  to point-source studies. The \Al\ emission is a good illustration: 
the only available \Al\ map to date has been released, more than fifteen years ago, thanks to the \COMPTEL\ instrument. 
However, at the present time, the \SPI\ spectrometer aboard the \INTEGRAL\
 mission offers a unique opportunity to enrich  this first result. In this paper, 
\(2 \times 10^8\) s of data  accumulated between  2003 and 2013 are used to perform 
 a dedicated analysis, aiming to deeply investigate the  spatial morphology of the \Al\ emission. 
The data are first compared  with  several sky maps based on  observations at various wavelengths to 
model the \Al\ distribution throughout the Galaxy. For most of the distribution models, the inner Galaxy flux is  
compatible with a value of 3.3$\times$ 10$^{-4}$ \phf while the preferred template maps  correspond to   young stellar components 
such as core-collapse supernovae, Wolf-Rayet and massive AGB stars. To get more details about this emission, an  
image reconstruction is performed using an algorithm based on the maximum-entropy method. 
In addition to the inner Galaxy emission, several excesses  suggest that some sites of emission are linked to 
the  spiral arms structure. 
Lastly, an estimation of the \Fe\ line flux,  assuming a spatial distribution similar to
 \Al\ line emission, results in a \Fe\ to \Al\ ratio 
around 0.14, which  agrees with the most recent studies and with the SN explosion model predictions. 
\end{abstract}   
 

\section{Introduction}\label{sec1}
The 1.809 MeV line emission associated with the \Al\ decay is one of the most intense gamma-ray lines observed in our 
Galaxy. It was first detected by the Ge spectrometer on the \HEAOC\ spacecraft 
\citep{Mahoney84}. However, to date, only the \COMPTEL\footnote{The \COMPTEL\ 
telescope \citep{Schoenfelder93} was studying the band 1-30 MeV, over a wide 
field-of-view of about 1 steradians. Around 1.8 MeV, the energy resolution is about 140 
keV and  angular resolution $\sim$4$^\circ$.}  
Compton telescope aboard the \textit{Compton 
Gamma Ray Observatory}  has mapped the \Al\ during its nine years survey. 
The emission has been found  mainly distributed along the Galactic plane and  supports a 
massive-stars origin \citep{Diehl95, Oberlack97, Knodlseder99, Pluschke01}. 
In addition, the early \COMPTEL\ skymaps suggest a number of marginally significant spots, some of them being potentially  associated 
 with the Galactic  spiral arms structure \citep{Chen96}.
However, most of these features remain compatible with statistical noise in the data \citep{Knodlseder99b}. \\ 
These  \COMPTEL\  maps have been used as a basis to fix the spatial morphology of the \Al\ line emission 
for  subsequent works related to the  spectral analyzes. 
Among them, detailed  studies in the inner Galaxy and extended regions 
along the Galactic plane  made with \INTEGRALSPI\  indicate that the intrinsic line width is less than 1.3 keV 
and that  line position shifts along the plane  corresponding to the rotation of our Galaxy,
confirming at least a partial  association of the \Al\ emission with the spiral arms
\citep[][and references therein]{Diehl06, Wang09, Kretschmer13}.\\
A related topic is the \Fe\ emission, more precisely the isotope lines at 1.173 and 1.333 MeV
released when it decays into $^{60}$Co and $^{60}$Ni, for the final stage of
 massive star evolution. 
Because of its weakness, the \Fe\ radiation has been detected from our Galaxy with only two 
instruments, \RHESSI\ \citep{Smith04} and \SPI\  \citep{Harris05, Wang07}. The  \Fe\ to 
\Al\ flux ratio is found to be  about 0.15 by both instruments. \\
In this paper,  10 years of INTEGRAL observation are used  to examine the spatial 
morphology of the \Al\ line, through direct sky-imaging and  sky distribution model comparison. 
We will emphasize on the data analysis, especially the instrumental background modeling 
issue, a key point for both \COMPTEL\ and \SPI\ instruments. We have developed several methods, and try to 
derive reliable conclusions, independent  of any specific (sky or background) model.\\
In the following sections, we review the main characteristics of the instrument and 
data-set and discuss the  basic principles of the developed methods  in 
Section~\ref{sec2}. We then present the results (Section ~\ref{sec3}) on the global morphology  of the
  \Al\ line emission, but also on specific regions suspected to harbor  \Al\ 
 progenitors,  before discussing the outcomes  in Section~\ref{sec4}.
%
\section{Data and analysis method}\label{sec2}
%
\subsection{Instrument and observations}
The  \textit{International Gamma-Ray Astrophysics Laboratory} (\INTEGRAL) observatory was launched from 
Ba\"{\i}konour, Kazakhstan, on 2002 October 17. \\
The on-board \SPI\ spectrometer  is equipped with an imaging system sensitive both to 
point-sources and extended/diffuse emission. It consists of a coded mask associated with 
a 19 Ge detector camera. This leads to a  spatial resolution of $\sim$2.6$^{\circ}$ over a 
field-of-view of 30$^{\circ}$   \citep{Vedrenne03}. The instrument's in-flight 
performance is described in \citet{Roques03}. Due to its non-conventional coded mask 
imaging system, the imaging capability relies also on a specific  observational strategy 
based on a dithering procedure \citep{Jensen03}, 
where the direction of pointing of each exposure is 
shifted from the previous one by \(\sim\)2.2$^{\circ}$. A revolution lasts 3 days, the 
time that the spacecraft performs a large eccentric orbit, but half of a day of data is unusable because of the radiation belts crossing. It contains generally about 100 exposures 
lasting approximately 45 minutes each.\\
The present analysis is based on public data recorded with the \SPI\ instrument from  
revolution 44 (2003 February 23) to revolution 1287 (2013 April 28). 
Around 1 MeV, high-energy particles saturate the electronics and can generate false 
triggers. Nonetheless, it is possible to analyze the signal in this energy range 
thanks to another electronic chain (via 
Pulse Shape Discriminators or PSDs) not affected by the saturation problem. The procedure 
is explained in \citet{JR09}.
We use the events which trigger only one detector (single-events). Note that the events which hit successively
two or more detectors (multiple-events, representing \(\sim\)25$\%$ of the photons in this energy band) are not used in this work.
After  excluding data contaminated by solar flares or the radiation belts, it results in 
about 77\,000 exposures and 2$\times$10$^8$ s observation livetime.\\ 
We performed our analyzes of the \Al\ line emission in the 1805-1813 keV band,  to  take into account 
the Germanium energy resolution (FWHM of 2.9 keV at 1764 keV), including its degradation  between two consecutive annealings ($\sim 5\%$). 
At these energies, the gain calibration (performed orbit-wise) accuracy is better than 
$\pm$0.01 keV. 
%
\subsection{Data modeling}\label{Sec2:model}
%
The signal recorded by the \SPI\ camera on the 19 Ge detectors is composed of 
contributions from each source (point-like or extended) present in the field-of-view, convolved by 
the instrument aperture, plus the background. For extended/diffuse sources, we assume 
that their spatial distributions are given by an analytical function or an emission map (Section~\ref{sec_templates})
whose intensities are to be determined.  
For a given energy band and for $N_s$ sources located in the field of view, the data $n_{dp}$ obtained during an  
exposure (pointing) p for the detector d can be expressed by the relation:
\begin{equation}
n_{dp}=\sum_{j=1}^{N_s} R_{dp,j} s_{j} + b_{dp}+\epsilon_{dp}
\end{equation}
where $s_{j}$ is the  intensity of source j.
$R_{dp,j}$ is the response of the instrument to source j, $b_{dp}$ the background,
and $\epsilon_{dp}$ the statistical noise (both for
exposure p and detector d). We assumed that there are $N_p$ exposures and $N_d$ detectors.\\
At the energies considered in this paper (E $>$ 1 MeV), point-source  emissions are weak and 
stable in time within the measurement uncertainties. Concerning the extended/diffuse sources, they are 
not expected to vary. 
Thus, for a given set of $N_p$ exposures and $N_d$ detectors, the system of equations, as formulated above, 
requires  $N_p \times N_d$ equations to solve for  $N_s + N_p \times N_d$ unknowns. 
Hence, it is mandatory to reduce the number of unknowns.  
We will see below that the observed properties of the background  
allow us to strongly decrease the corresponding number of free parameters.
Finally, to determine the sky model parameters, we adjust the data through a 
multi-component fitting algorithm, based on the maximum likelihood  statistics. 
Expected counts are obtained by convolving a sky model with the instrument response and then adding the background model. The resulting distribution is compared to the recorded data with free normalizations of both components.
We used Poisson's statistics 
to evaluate the adequacy of the various  sky models to the data.  The core algorithm developed to handle 
such a large system is described in \citet{Bouchet13a}.
%
\subsection{Modeling of the \Al\ spatial distribution} \label{sec_templates}
\begin{figure*}[!ht]
\centering
\plotone{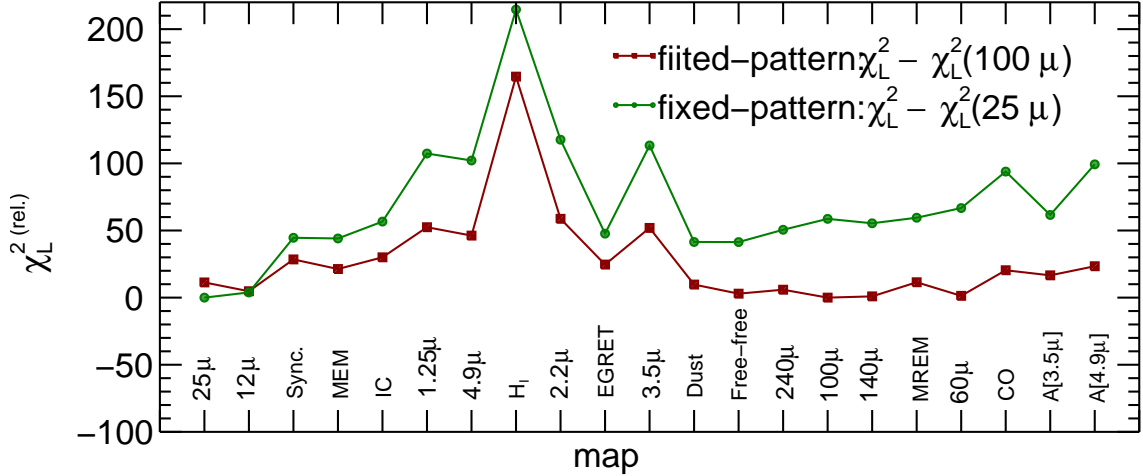}
\caption{
The relative chi-square variation ($\chi^{2 (rel.)}_L)$ versus assumed  template to model the distribution of the \Al\ line. $\chi^{2 (rel.)}_L)$  is the $\chi^2_L$  from which the value of the best fitted template is subtracted.
The best template is the 100$\mu$ template for the fitted-pattern method ($\chi^2_L$=1258778.1 for 1261797 dof) and the 25$\mu$ template for the fixed-pattern
method ($\chi^2_L$=1266968.4 for 1262208 dof).
Terms: sync., dust and free-free, are abbreviations for 53 GHz synchrotron, dust and  
free-free maps described in Table~\ref{table:tracers}. 
MEM and MREM indicate the \COMPTEL\ maps, A[3.5\(\mu\)]  
and A[4.9\(\mu\)]corrected NIR extinction map. 
The red curve is for the fitted-pattern and green for fixed-pattern method. 
The maps are ordered following their contrast defined as the ratio of the flux 
enclosed in the  region $\vert l \vert < 150^{\circ}$, $\vert b \vert < 
15^{\circ}$  to the total flux.
}
\label{fig:rmlr}
\end{figure*}

%
A way to estimate the \Al\ emission spatial distribution over the Galaxy (i.e. the corresponding $s_{j}$ term in Equation 1) 
is to represent it with some templates. 
In this study, we have used maps listed in Table~\ref{table:tracers}, similarly to \citet{Knodlseder99} for the \COMPTEL\ data.
These maps, although the list is not exhaustive, emphasize some large-scale structures of the sky observed 
at particular 
wavelengths and associated to  specific emission mechanisms that we may relate to the \Al\ emission.\\
Specific treatments have been applied to some of them:
The A[${3.5\mu}$] and \Ared4.9\ maps are the NIR 3.5 and 4.9$\mu$  maps corrected for reddening using NIR 1.25$\mu$ map and averaging emission at latitude 
$|b| > 40^\circ$ to estimate the zero level emission as explained in \citet[][and references therein]{Krivonos07}.
Note that, with this procedure, the extra-galactic component has been removed, but the resulting maps are not 
expected to have an  accuracy better than \(\sim 10\%\).
For the CO \citep{Dame01} and the EGRET ($>$ 100 MeV) (provided by the NASA/Goddard Space Flight Center)  maps, we apply the pre-treatment 
detailed in \citet[][and references therein]{Knodlseder99}.
Hence, the peak emission around the Galactic Center of the CO map has been removed, while  point-sources from the second EGRET catalog have been  subtracted  from the \EGRET\ map, together with
 an isotropic intensity of 1.5$\times$ 10$^{-5}$ ph\ cm$^{-2}$\ s$^{-1}$\ sr$^{-1}$\ to take into account the cosmic diffuse background radiation.\\
We have introduced a parameter to quantify the difference between all these maps, the 
contrast, which we defined as the ratio of the flux contained in the region $\vert l \vert \leq 
150^{\circ}$, $\vert b \vert \leq 15^{\circ}$ to the total flux 
enclosed in  $\vert l \vert \leq 180^{\circ}$, $\vert b \vert \leq 90^{\circ}$. 
This ratio varies from  0.4, for the 25$\mu$ MIR map, to  almost 1, for the \Ared4.9 map. 
For reference, the 
 EGRET map, which traces the  interstellar gas/cosmic-ray emission, has a 
contrast value of 0.7. The maps up to EGRET (left part of the x-axis of~Figures~\ref{fig:rmlr} and~\ref{fig:flux})
 are said to have low-contrast and those above (right part) high-contrast.
%
\subsection{Background determination}\label{sec_background}
The instrumental background corresponds to a more or less  isotropic component due to 
particles hitting the telescope or created inside  its structure. It is the main 
contributor to the flux recorded by the detector plane and represents a key issue since the   
signal-to-noise ratios considered in this study are below 1$\%$.\\
However, we have to note  that the background term in Equation (1), formally 
consisting of $M=N_d \times N_p$ values (one per detector and per pointing), can be rewritten 
as:
\begin{equation}
b_{dp} = a_p \times u_d\times t_{dp}
\end{equation}
Here \(u\) is a vector of $N_d$ elements, representing the "uniformity map" of the detector plane (background  pattern),
$t_{dp}$  the effective observation time for detector d and pointing p.
The evolution of the background intensity is traced with {$a_p$},  a scalar normalization coefficient per pointing.
Assuming that \(u\) is determined elsewhere, 
the number of unknowns (free parameters) in the  set of $N_p \times N_d$ equations is 
reduced to $N_s + N_p$ (for a background intensity varying on the 
exposure timescale).
However, the background intensity ($a_p$)  does not necessarily vary so rapidly and the number of related 
background parameters could be still decreased, according to the actual background evolution. \\
We detail below both aspects (detector pattern and timescale evolution) of the background determination.
%
\subsubsection{Background intensity variations}\label{sec_intensity}
\begin{figure*}[!ht]
\plotone{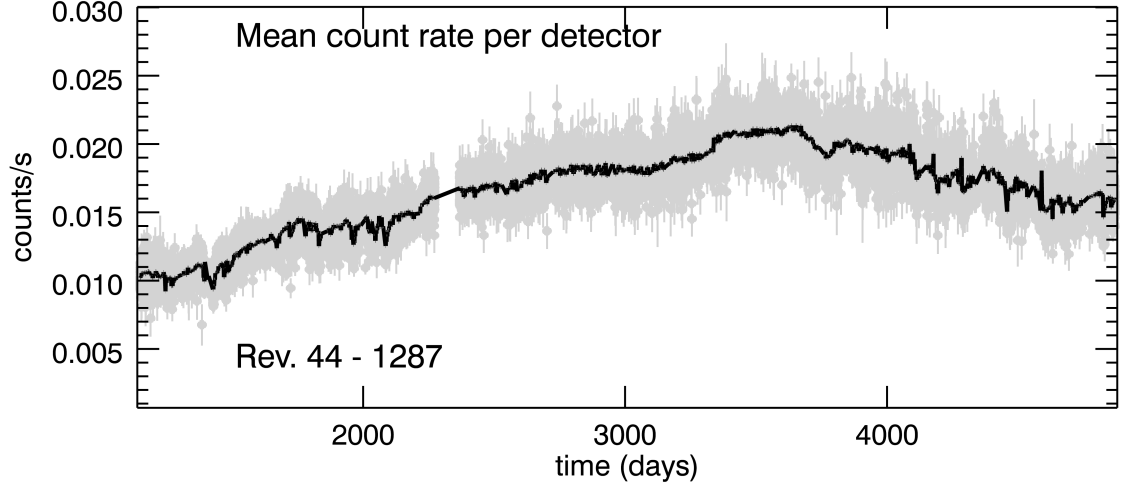} 
\caption{
Mean count rates recorded on the camera per detector  versus  time
 in the 1805-1813 keV energy band. 
In gray, the count rates and associated error bars (1 point per exposure). In black, the count rates averaged per revolution.
The total number of exposures is 76\,789. 
The hole around days 2500 corresponds to exposures with no PSD information 
(Section~\ref{sec2}).}
\label{fig:bkgrate}
\end{figure*}

\begin{figure*}[!ht]
\plotone{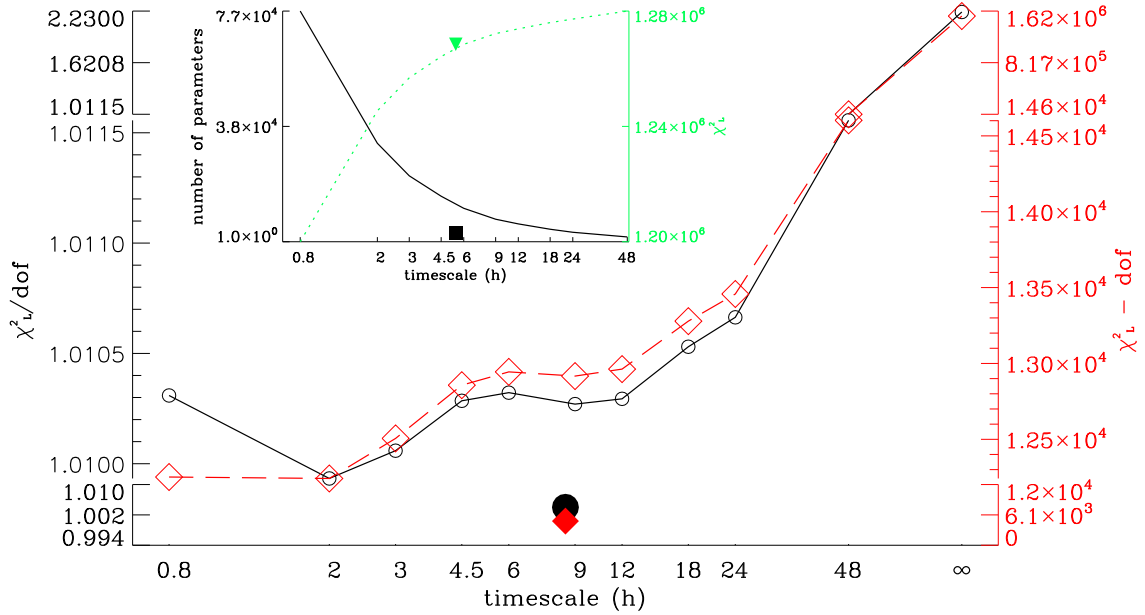}
\caption{
A  sky model, consisting of a synthetic map (here the 60$\mu$ map, Section~\ref{maps_template_txt}), 
plus a few sources, is fitted to the 1805-1813 keV  data (fixed-pattern method). 
The evolution of the reduced ($\chi^2_L/dof$) (solid-line, open black-circles) 
and the 
$\chi^2_L - dof$ (dashed-line,  open red-diamonds) are shown as a function of the assumed 
background timescale (in hours).
In insert  are displayed the $\chi^2_L$ (dotted-green line) and the number of parameters to be 
determined (solid black line). 
The filled black-circle, red-diamond and (in insert) green triangle and black-square 
show respectively the $\chi^2_L/dof$, the $\chi^2_L - dof$ , the  $\chi^2_L$ and the  number of parameters 
obtained with the background segmentation method.
The infinity symbol ($\infty$) on the x-axis corresponds to a constant background for the whole dataset.}
\label{fig:bkgtimescale}
\end{figure*}

\begin{figure*}[!ht] 
\plotone{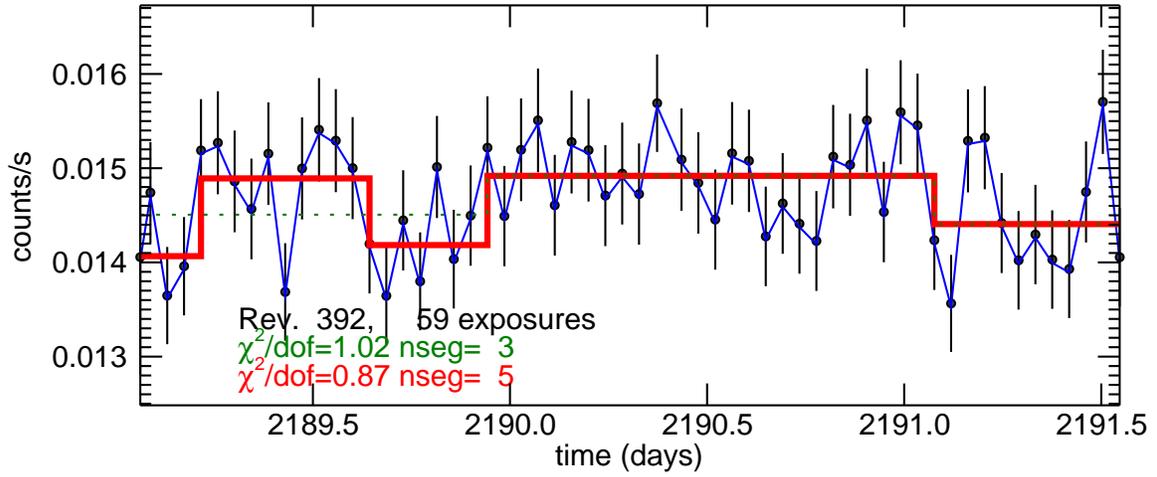}
\caption{Background intensity evolution in revolution 392.
The background is modeled  with 3 time-segments 
(dotted purple-line) to give a  $\chi^2/dof$ of 1.02 or with 5 time-segments (red line) 
for a $\chi^2/dof$ of 0.87.}
\label{fig:bkg_variable}
\end{figure*}

%
Figure~\ref{fig:bkgrate} shows the mean count rate (per detector) evolution with time. 
In our standard analysis, we generally   
assume that the background varies with a fixed timescale. We have tested several timescales 
from one exposure ($\sim$ 2-3 ks) to one revolution ($\sim$ 2-3 days) and compared the 
results from a statistical point of view. Figure~\ref{fig:bkgtimescale} shows the evolution 
of several indicators as a function of the chosen timescale. The "reduced chi-square" is defined as 
$\chi^2_L/dof$, where $\chi^2_L$  stands for  the likelihood equivalent chi-square\footnote{
To obtain a \(\chi^2\) equivalent statistics for  Poisson-distributed data, 
\citet{Mighell99} proposes the usage of the maximum likelihood ratio
statistic 
$\chi^2_L \equiv 2 \times \sum_{i=1}^{M} e_i-n_i+n_i \times ln(n_i/e_i)$,
\(M\) is the number of data points and \(e_i\) is the model of the data \(n_i\).
This is the Cash statistic  \citep{Cash79} for  Poisson-distributed data (low counts); 
$C \equiv -2 \times ln(L) = 2 \times \sum_{i=1}^{M} e_i-n_i \times ln(e_i)$, where \(L\) is the likelihood function, within
the constant term $\sum_{i=1}^{M} n_i-n_i \times ln(n_i)$.
}
and \(dof\) for the degrees of freedom.\\  
The  reduced chi-square  (or similarly $\chi^2_L - dof$) is rather stable for timescales less than
 12 hours; above, its value increases very 
quickly. As we aim to have as few as possible free parameters, in order to get a
 better conditioned system of equations, we can consider that   
a timescale of $\sim$ 9 hours is a reasonable trade-off. In comparison, for the 505-516 keV band, 
we have  concluded in \citet{Bouchet10} that the best quality and robust results are obtained 
with a value of $\sim$ 6 hours. Note that the behavior of the above quantities does not 
depend on the assumed synthetic map nor on the pattern determination method 
(Section~\ref{sec_uniformity}).\\ 
However, to impose segments of fixed duration is convenient and easy, but  not 
necessarily the best way to proceed since the background intensity can vary with 
various timescales along the mission. 
 We have therefore applied a segmentation algorithm developed to determine the variability
 timescale of the sources \citep{Bouchet13b} to the background signal. 
We first subtracted the source contribution from the total counts. A 
rough approximation of the sky signal is enough for this 
purpose (its contribution to the total counts is weak). 
Then, for each exposure, counts are summed over all detectors, 
and for each revolution, the corresponding time-series (up to one hundred of exposures) is segmented. 
The number of segments and their lengths are adjusted  in order to obtain 
the minimal number of segments  ensuring $\chi^2$/dof $\leq$ 1. 
Figure~\ref{fig:bkg_variable} illustrates the case  where the   
background intensity  varies with different  timescales along the revolution.\\
The background intensity has been found  stable (fit with one segment) during  477 over
1081 revolutions. 
Finally,  the total number of segments has been significantly reduced since 
only $\sim$3000 segments are required  to describe  the whole dataset, 
instead of the  $\sim$12\,000 segments used when fixing a $\sim$6 hours timescale. 
To compare qualitatively these results with those obtained with fixed timescales,
 we plot in Figure~\ref{fig:bkgtimescale} the statistical quantities 
 corresponding to the background segmentation method with isolated symbols (arbitrary abscissa
 since no fixed timescale).  
All of them point toward an improvement of the fit quality when the background variability is 
determined with a flexible timescale.
Finally,  these (quasi model-independent) time segments have been used to describe the background 
evolution in our subsequent analyzes.
%
\subsubsection{"Uniformity map"}\label{sec_uniformity}
%
The uniformity map or background pattern (\(u\) in Equation 2) can be fixed by hand before the 
fitting procedure by using ``empty-field'' observations, thought to contain no source 
signal. 
The dedicated \SPI\  ``empty-field'' observations are rare, but the exposures whose 
pointing latitude direction satisfies $|b| > 30^\circ$ constitute a good approximation  
since they contain only weak contributions from sources, at the energies considered here. 
They amount to 20$\%$ of the observations and have been  used  for building a set of background 
templates for different periods along the mission (about one per 6 months).\\
To quantify the properties of the background pattern for each 
  period, we define the vector \(u\) as 
 \begin{equation}
u(d)=\frac{\sum_{p} d_p(d)} {\sum_{p} t_p(d)}~\mathrm{for~exposure~p~satisfying~|b| \ge 30^\circ}
\end{equation}\
However, in the case of diffuse emission, the high-latitude exposure fields may contain some signal and the background 
pattern deduced from them may be "blurred". More precisely, the high-latitude  $|b| > 30^\circ$ regions 
contain $\sim$ 30$\%$ to 40$\%$ of the total emission  for the "low-contrast" maps (12$\mu$, 25$\mu$ and synchrotron) 
while this ratio is less than 10$\%$ for "high-contrast" (EGRET to \Ared4.9 maps) ones. Consequently, if the true 
emission distribution approaches the 25$\mu$ map, the detector pattern will be more affected by the high-latitude 
emission than if the true emission distribution follows the \Ared4.9 map. Note that the effect on the measured 
source flux is not predictable since the fit procedure adjusts background and source normalizations: a ``bad'' 
background pattern may imply  an underestimation or an overestimation of the source flux.\\
In addition, for the \Al\ study, we have to keep in mind that the side shields do not stop 100\% 
of  photons. 
This means that any uniformity map or background pattern 
will contain  also some diffuse emission signal passing through the shield.\\ 
We have investigated another approach to estimate the detector uniformity pattern, by fitting
 it during the convergence procedure. We have to note, in this case, that a prerequisite is to 
 have a sufficient knowledge of the source contribution and that the  background  model relies 
 on more free parameters. 
 With this alternative method, the 
$\chi^2_L$  decreases by $\Delta \chi^2_L \sim$8200   for $\Delta$dof=411 additional 
parameters (2973 parameters are used to describe the background). 
However, the improvement of the $\chi^2_L$ criterion  is not reliable enough since the recovered 
source signal becomes background dependent and can be altered to an extent which is difficult to estimate. \\ 
 The two pattern determination methods are subsequently referred to  as fixed-pattern and fitted-pattern ones, 
and we have systematically compared the results obtained with each of them.
  
\begin{figure*}[!ht]
\plotone{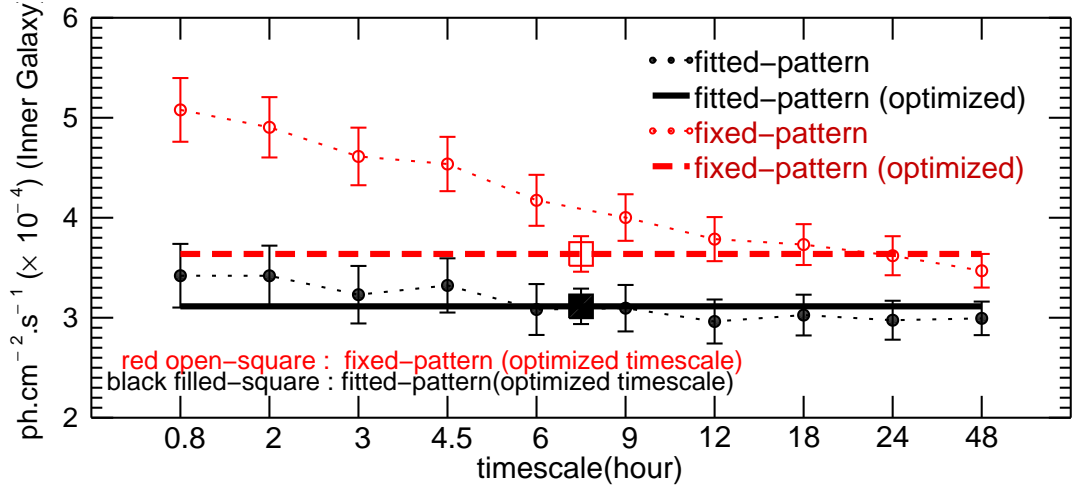}
\caption{
Flux of the sky obtained  with the fixed-pattern (open red-circles) and 
fitted-pattern (filled black-circles) background method as a function of the 
assumed background variation timescale (here using a 60$\mu$ map). 
The values obtained with the optimized number of background parameters 
are shown with the squares (open for fixed and filled for fitted-patter).
}
\label{fig:bkg_fluxes}
\end{figure*}

Figure~\ref{fig:bkg_fluxes} presents the \Al\ line flux obtained in the inner Galaxy ($\vert l \vert \leq 30^{\circ}$, $\vert b \vert \leq 10^{\circ}$) for both pattern 
determination methods and several background variability timescales 
with the same model for the \Al\ distribution (here the 60$\mu$ template). 
With the fixed-pattern method, the recovered flux depends 
on the assumed background timescale, especially for background timescales below 6 hours, 
while its value remains unaffected when the background  pattern is adjusted.
 It is worth noting that the same analysis has been applied to the 
annihilation radiation, at 511 keV. Due
to a 
better statistic and possibly a relatively less emission at high-latitudes, 
the fluxes obtained with the two pattern determination methods are perfectly 
compatible for each of the sky components, whatever the assumed background variation 
time-scales.\\ 
%
\subsection {Imaging the sky}\label{sec_imaging}
Although the  model-fitting process provides the best quantitative information about the global emission and is 
better suited to determine its level of confidence, a direct ``imaging'' algorithm provides 
more qualitative information such as the position of potential emitting sources and their 
extent as well as some basic, but model independent, characteristics of the 
emission morphology.\\
To build an image, the sky is divided into small areas or pixels, the flux \(f\)  in each 
pixel is to be determined. The linear model of the data derived from  Equation (1) is put in a more 
synthetic form through
\begin{equation}
n = Rf + b + \epsilon
\end{equation}
where \(n\) represents the data, \(b\) the background, \(R\) the response of the instrument as described in 
Section~\ref{Sec2:model} and \(\epsilon\)  the  statistical noise.  \(n\), \(b\)  and \(\epsilon\) are vectors of 
length  \(M\) (the number of data points), \(f\) a vector of length \(N\) (the number 
of pixels in the sky) and \(R\) a matrix  of size \(M\) by \(N\). 
The  statistical noise is assumed to 
follow a  Gaussian distribution with a variance \(\sigma^2\) and null mean.\\
However, when the number of pixels is large, the system of equations  becomes ill-conditioned;  
the direct least-square solution is not always reliable and generally poorly 
informative. 
To select an informative and particular solution, the most common technique consists in 
including a regularization term in the above system, in addition to the least square 
constraint. This solution, although biased by the choice of the regularization criterion,  improves the conditioning of the system. A 
compromise between the goodness of the fit, quantified by \(\chi^2\),
and the regularization, say 
\(H\), is found by maximizing the function
\begin{equation}
Q(f)=\alpha H(f)-\frac{1}{2} \chi^2(f)
\end{equation}
where \(\alpha\) is a parameter, which determines the degree of smoothing of the 
solution. 
\\
We choose to use one of the most popular regularization operators: the entropy function. This 
method is known as the Maximum-Entropy Method or MEM \citep[][and references therein]{Gull89}. 
For an assumed positive additive distributions,
\begin{equation}
H(f)=\sum_{i=1}^N f_i -m_i -f_i log \frac{f_i}{m_i}
\end{equation}
where \(m_i\) is the default initial value assigned to the pixel \(i\). 
Note that \(f_i\) and \(m_i\) are positive quantities.
MEM has been proved to be 
a successful technique for astronomical image reconstruction \citep{Skilling81}.
The algorithm is described in Apppendix~\ref{appendixA}.
However, despite 
its capabilities,  it suffers from several shortcomings: among them, the difficulty of selecting 
the appropriate entropy function distribution and the default pixel values. In addition, 
the possible correlations between pixels are not taken into account properly. 
One way to remedy  this problem is described in Section~\ref{mem_practical}.
%
\subsubsection {Including the background intensity determination}\label{sec_imaging_bkg}
%
In our application, the  instrumental background could be fixed through 
the modeling method  presented in Section~\ref{sec_background}.
However, such a modeling, even though  being accurate enough for the  model-fitting procedure,
may be a source of biases for the image reconstruction, by preventing the appearance or by 
exaggerating the significance of some structures potentially present in the image. 
Consequently, we prefer to have the possibility to refine the parameters linked to the 
background intensity during the image reconstruction process.
The instrumental background variation timescale is fixed through 
the modeling method  presented in Section~\ref{sec_background}.
We search for both the solution vector \(f\) (length \( N \)) and the background intensities, \(b\) (length \(N_b\)).
We assume that the background parameters are precisely determined/constrained (small 
error bars) compared to sky pixels, hence they do not require any smoothing.
Note that 
the  "cross-talk" effects between sky pixels and background parameters are reduced since 
the number of parameters to describe the background is nearly minimum 
(Section~\ref{sec_intensity}).
%
\subsubsection {Pixel correlation}\label{mem_practical}
As mentioned above, the possible correlations between the pixels of the image 
(the sky emission should vary smoothly from one pixel to the next ones) 
are not taken 
into account properly with the entropy function. It is possible to introduce artificially such a correlation 
 by using an Intrinsic Correlation Function, ICF, \citep{Gull89}. The function to maximize becomes:
\begin{equation}
Q(f)= \alpha H(f)-\frac{1}{2} \chi^2(f_{ICF}) ~\mathrm{where}~f_{ICF} \equiv ICF(f)
\end{equation}
For our work, we choose  a radial Gaussian function. Such
 an ICF takes part largely to the solution regularization by ensuring its 
smoothness. 
\subsubsection {Construction of a sky image}\label{code_practical}
In practice, the sky is divided into pixels of equal area, following the Healpix (Hierarchical 
Equal Area isoLatitude Pixelization of a sphere) scheme \citep{Gorski05}, and initialized
with a uniform default value. 
In a first step, we consider a very low-resolution map (48 pixels of $\sim 29^\circ$ resolution)
 and select the value of \(\alpha\) such that the reconstructed flux in the 
inner Galaxy is  comparable to the value obtained with the model-fitting procedure.
The size of this problem is relatively small and a N-dimensional search-algorithm is 
used to maximize the function \(Q\) ($\sim$3000 parameters). The calculation of the first-order solution 
(image and background parameters) is based on a classical Newton type optimization algorithm  
with a positivity constraint on the solution. 
During this first stage, the background pattern is fixed. The resulting image is used as a template to compute an improved pattern which is then fixed for the rest of the process.
We then use this solution as a starting point 
(for an initial guess of the background parameters, the image is assumed to be flat)
to build a high-resolution image (49\,152 pixels of 0.9$^\circ$ resolution) through an 
 algorithm similar to that proposed by \citet{SB84}. 
 The parameter \(\alpha\) and  
the uniform sky default model  are computed at 
each iteration so that the solution is ensured to follow the optimum MEM trajectory (See Appendix~\ref{appendixA}). 
At the end, the final solution satisfies $\chi^2/dof \simeq 1$ 
and fulfills an additional test on the degree of non-parallelism between the gradient of \(C\) and \(H\) (see Appendix~\ref{appendixA}). 
Note that,
the greater the number of iterations is, the lower the chi-square value but the image is more ``spiky''.
The above mentioned stopping criteria results probably in a "spiky'', but also more objective image \citep{SB84}.
%
\section {Results}\label{sec3}
%
\subsection{Imaging}
\begin{figure*}[!ht]
\centering
\includegraphics[width=12cm]{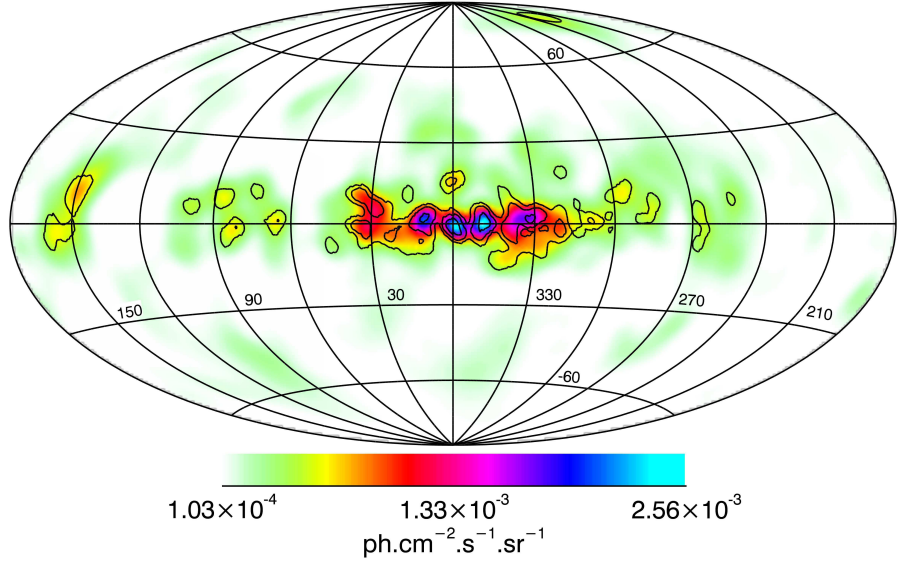}
\caption{Image of the \Al\ line (1805-1813 keV). 
The image is built with a resolution 
 (ICF FWHM)  of 6$^\circ$. For this image reconstruction, the background pattern is adjusted as explained in Section~\ref{code_practical}.
The contours are extracted from the 3$^\circ$ resolution image.
In units of $\times 10^{-3}$ ph\ cm$^{-2}$\ s$^{-1}$\ sr$^{-1}$, they correspond to
0.54, 1.1, and 2.7. 
Identified regions, from left to right: Perseus region (105$ ^\circ$ $\leq$ l $\leq$ 170$ ^\circ$) (Taurus clouds), 
Cygnus/Cepheus region (75$ ^\circ$ $\leq$ l $\leq$ 100$^\circ$), 
the inner Galaxy (-30$ ^\circ$ $\leq$ l $\leq$30$ ^\circ$, -10$ ^\circ$ $\leq$ b $\leq$10$ ^\circ$),   
Carina (l=286$ ^\circ$, b=1$ ^\circ$) and 
the  Vela region (260$ ^\circ$$\leq$ l $\leq$ 270$ ^\circ$).
At mid-latitude, the Sco/Cen region (300$ ^\circ$ $\leq$ l $\leq$ 360$ ^\circ$, 8 $ ^\circ$$\leq$ b $\leq$ 30$ ^\circ$). }
\label{fig:image_alu}
\end{figure*}

%
The image displayed in Figure~\ref{fig:image_alu}  
indicates that the emission is essentially confined in the inner Galaxy region 
($\vert l \vert \leq 30^{\circ}$, $\vert b \vert \leq 10^{\circ}$), 
with a  flux in the inner Galaxy of 3.5$\times 10^{-4}$\phf.
Note that the morphology is drastically different from the electron-positron annihilation line emission one, 
which is essentially concentrated in the bulge region
\citep{Weidenspointner08, Bouchet10, Churazov10}. \\
 Beyond the morphology of the diffuse emission, our interest has been piqued by 
several spots visible in the image. 
Simulations of image reconstruction (Appendix~\ref{appendixB}) show that structures with a peak intensity 
at $\sim$one-tenth of the image maximum intensity are probably due to residual statistical noise. Above this level, structures are worth being considered
carefully.
Indeed, some of them have  positions compatible with sources 
detected or studied during recent \SPI\ investigations (See section~\ref{region_excesses}).
For instance, the image reveals several excesses in the  Cygnus region (72$^\circ < l < 96^\circ$,  -7$^\circ < b <7^\circ$).
A significant one is located at a position compatible with Cyg OB2 cluster (81$^\circ$,-1$^\circ$) 
as reported by \citet{Martin09}.
In the Sco-Cen region (328$^\circ  < l < 355 ^\circ$, 8$^\circ < b < 30^\circ$), the strongest excess is localized at (l, b) $\approx
$(360$^\circ$, 16$^\circ$), having a spatial extent with a radius of 5$^\circ$. 
Some structures appear also around the Carina and Vela regions and two others, more
 extended, in the Taurus/Anticenter region (105$^\circ <l <170^\circ$,
 -15$^\circ <b <20^\circ$). 
Finally, an additional spot detected above 3$\sigma$ is worth investigating. 
However, a model-fitting analysis is required to extract more quantitative information on the  above mentioned structures (Section~\ref{region_excesses}).
%
\subsection{Testing template maps} \label{maps_template_txt}
\begin{figure*}[!ht]
\centering
\plotone{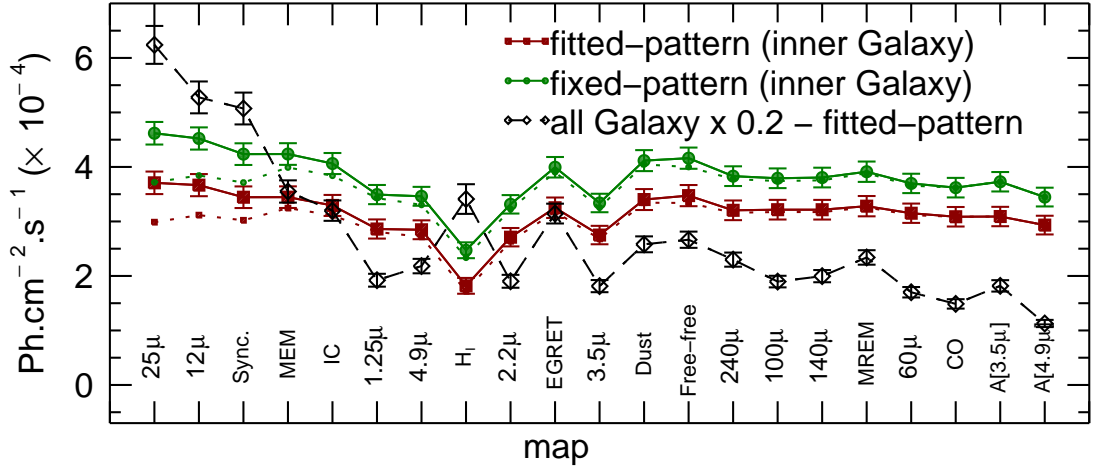}
\caption{Flux in the inner Galaxy as a function of the map used to model the distribution 
of the \Al\ line. The labels are the same as in Figure~\ref{fig:rmlr}. 
The dashed black-curve is the total flux in the Galaxy (fitted-pattern), scaled by a 
factor 0.2.
The dotted lines (red for fitted-pattern and green for fixed-pattern) are 
the fluxes obtained, if an isotropic emission (possibly extra-galactic) 
estimated by using the map emission at $\vert b \vert >$40$^\circ$, is subtracted from 
each template.
The labels are the same as in Figure~\ref{fig:rmlr}. 
}
\label{fig:flux}
\end{figure*}

%
This analysis is similar to that done with the \COMPTEL\ data by \citet{Knodlseder99}, and consists of using template maps to 
model the spatial distribution of the  \Al\ emission through the Galaxy. 
In this case, only the  intensity normalization is adjusted (1 parameter) in addition to the background parameters. 
In practice, we fit the data  with a combination of one of the maps listed in 
Table~\ref{table:tracers} plus a background  model. The template preparation is detailed in Section~\ref{sec_templates}.\\
To qualify the detection significance of the \Al\ emission for a given template,
we adjust two models to the data: the first one contains only the background while the second contains the background plus the tested template 
map.  Quantitatively, the improvement of the likelihood is 512 for 1 additional parameter for the 25$\mu$ map (best model) and the background determined with
the fixed-pattern method.
The \Al\ flux detection significance is 
\(\sim 23 \sigma\), and the associated reduced-chi-square is \(\chi^2_L /dof =1.0038 \). 
Indeed, each template leads to a significant flux detection. The worse case, H$_I$ map, has a significance of \(\sim 17 \sigma\).\\
The same analysis with the fitted-pattern background determination gives an 
improvement of the likelihood of 312 (for the 100$\mu$ map). The \Al\ flux detection significance is \(\sim 18 \sigma\) and \(\chi^2_L /dof =0.9973 \).\\
Figure~\ref{fig:rmlr} displays the Maximum Likelihood Ratio (MLR) obtained with both 
pattern determination methods (fixed and fitted-pattern) for each of the tested maps 
(ordered by increasing contrast). 
Several maps give similar  results, leading to the conclusion
that star-related distributions (FIR and MIR 
maps,  dust and free-free distributions, all with rather high-contrast)  give
 a good description of the data, as 
expected from what we know about the \Al\ emission process. 
Similarly, 25$\mu$ and 12$\mu$ maps constitute good tracers, while presenting a low-contrast.   
In another hand, H$_ {I} $ and 53 GHz synchrotron maps can be excluded. 
In fact, we have seen (Figure~\ref{fig:image_alu})  that the emission is confined into the central part 
of the Galaxy and close to the disk. This explains why the H$_ {I} $ map does not provide a good fit 
to the data since it extends far in longitude and contains important emission 
at $\vert l \vert \ge 30^{\circ}$. 
We  can also point out that 
distributions built by imaging method (MEM or MREM) from \COMPTEL\ data  appear  as  good tracers 
of the \Al\ emission, as expected.\\
As shown in Figure~\ref{fig:flux}, the reconstructed flux attributed to the inner 
Galaxy  does not depend much on the template map and
  is  compatible with a value of $3.3 \times$ 10$^ {-4} $ \phf, 
in agreement with the previously reported values. 
We observe that the flux is systematically higher by $\sim$8 (low-contrast maps) to 20$\%$ (high-contrast maps) when the fixed-pattern background determination method is used, but it has no scientific implication.
However, the total flux integrated over the whole sky clearly depends on the contrast of the 
assumed model. The more the map is contrasted (from left to right on Figure~\ref{fig:flux}), the weaker is the 
reconstructed total flux. This is due to the fact that the recovered global intensity  
relies on the central parts of the image (both higher flux and signal to noise ratios) 
and that a contrasted map encompasses less flux 
in its external  parts than a flatter map. 
Moreover,  we are aware that low-contrast maps may suffer from significant ``cross-talk''
 between the low-spatial frequency structure (a kind of pedestal which mimics a flat low-surface brightness  
emission) and the (more or less uniform) background contribution.
\\
As an additional test, we have performed
correlations between the direct image with 6$^{\circ}$ resolution (e.g. ICF of 6${^\circ}$ FWHM) and each of the 
templates downgraded  to the same \(\sim\)6${^\circ}$ spatial 
resolution (except the DMR/COBE 53 Ghz one, which originally has 7${^\circ}$ resolution). Indeed, the linear correlation coefficient does not depend much
 on the template map and keeps a value  greater than 0.9 in latitude, and above 0.7 in longitude, except
  the H$_ {I}$ map (coefficient of 0.4 in longitude).\\
Finally, it appears that it is hard to firmly conclude about a unique solution. This 
reflects the difficulty of determining precisely such an extended weakly emitting structure 
and the similarity presented by most of the considered maps. 
However, we note that the $\chi^2$ curves follow the same evolution regardless 
the background  determination method and hence that the conclusions do not depend on it.
%
\subsection{Regions of potential excesses}\label{region_excesses}
%
To check quantitatively the significance of the most significant excesses and known \Al\ emitting regions, we have performed a more complete model-fitting analysis.
Note, however, that our analysis is not optimized for extended point-sources  since 
a map contribution is necessarily subtracted  at the position of the 
sources and, at worse, may make them disappear.\\
The sky model consists of one of the templates listed in Table~\ref{table:tracers}, 
to which is added a spatial model including the spots. For the sources already  detected or investigated at 1.8 MeV,  we have used the positions and spatial extensions provided by previous works (Vela \citep{Diehl95},  Cygnus region \citep{Martin09}, Sco-Cen \citep{Diehl10},   
Orion/Eridanus, \citep{Voss10} and Carina \citep{Voss12}, indicated in bold in Table~\ref{table:tracers}).
To investigate the additional spots not yet referenced as \Al\ emitters, 
the extent and location are based on the image analysis  using simple 
 models (point-source, Gaussian or disk). However, given their low-significance and the \SPI\ spatial resolution of 2.6 $^\circ$, their identification with known sources is just indicative.
We report in Table 2, the flux values obtained using the IC (low-contrast) and 
\Ared4.9\ (high-contrast) maps, as representative of the global results.
 \\ 
The  Cyg OB2 cluster (81$^\circ$,-1$^\circ$)
has a flux  of ($4.1 \pm 1.5$)\dixmoinscinq\,  similar to the value of 
($3.9 \pm 1.1$)\dixmoinscinq\ obtained by \citet{Martin09} and the ($3.7 \pm 1.1$)\dixmoinscinq\ obtained with \COMPTEL\ \citep{Pluschke00}.
We note also another excess  at (l,b)$\approx$(100$^\circ$,6$^\circ$) with a flux of (2.7 $\pm$ 1.1)\dixmoinscinq. \\
Strong emission has been reported in the Sco-Cen region by \citet{Diehl10} centered around 
(l, b) = (350$^\circ$, 20$^\circ$). At this position, we find a flux  ($4.1 \pm 1.6$)\dixmoinscinq\ when using an $H_I$ map to model the large scale \Al\ emission.
 However in our image, the local maximum in this region appears shifted to  (l, b) $\approx$(360$^\circ$, 16$^\circ$), and the source is less extended 
(disk radius of
5$^\circ$  instead of  10$^\circ$  as reported by  \citet{Diehl10}), while the flux remains at ($4.1 \pm 1.1$)\dixmoinscinq\ using again the $H_I$ map.
On the another hand, in both cases the flux clearly depends on the template map used and is less than 2$\sigma$ for most of the cases.\\
The Carina and Vela regions are marginally detected (2$\sigma$). For Carina, the measured flux of ($2.8 \pm 1.5$)\dixmoinscinq\ is comparable to the 
estimation of ($1.5 \pm 1.0$)\dixmoinscinq\ reported by \citet{Voss12} using \SPI\ data, and ($3.1 \pm 0.8$)\dixmoinscinq\ reported by \citet{Knodlseder96} using \COMPTEL\ data.
For Vela, the flux of ($3.3 \pm 1.8$)\dixmoinscinq\ is comparable to the value ($3.6 \pm 
1.2$)\dixmoinscinq\ reported by \citet{Diehl95b} using \COMPTEL\ data.
For the Orion-Eridanus area (180$^\circ < l <210^\circ$, -30$^\circ  <b <5^\circ$),
 we get an upper limit for the total flux  of 3\dixmoinscinq\ (1 $\sigma$) in agreement 
with  the value of (4.5 $\pm$ 2.1)\dixmoinscinq\, obtained by \citet{Voss10} from a synthesis model 
based on the Orion star populations. 
We also mention the   the 2$\sigma$ upper limit of 1.7\dixmoinscinq\ reported by  \citet{Oberlack95} for the Orion region using \COMPTEL\ instrument.
\\
Concerning the spots not yet reported, we first point out 
two extended structures in the Taurus/Anticenter region (105$^\circ <l <170^\circ$,
 -15$^\circ <b <20^\circ$) with fluxes of ($5 \pm 2$) and  ($8 \pm 3$)\dixmoinscinq.
Finally,  a spot worth mentioning since it is detected above 3 $\sigma$, it is  located 
at high-latitude (l,b)$\approx$(226$^\circ$76$^\circ$) with a flux of (7 $\pm$ 2)\dixmoinscinq.
We did not find any convincing potential counterpart,  but consider the stability of the attributed flux   against the 
assumed  \Al\ diffuse emission map as a good criterion  to assess that an excess is robust and reliable.\\
We did not notice any systematic effect in the source flux determination between the fixed and fitted background pattern adjustment method.
Note that adding all these excesses into the model of the sky improves  the likelihood only marginally ($\Delta \chi^2_L \sim 5$) compared to the case where they are neglected, and that    dedicated analyzes  must be conducted to refine individually each result.
%
\subsection{\Fe\ and \Fe\ to \Al\ ratio}
%
While the intensity of the \Fe\ isotope emissions at  1.173 and 1.333~MeV
contains important complementary information for the star evolution study, the weakness of the 
flux makes it impossible to derive  any constraint on the spatial distribution. 
The latter is, therefore, assumed to be the same as the \Al\ line one (Section~\ref{maps_template_txt}), 
which is physically reasonable since  \Al\ and \Fe\ are believed 
to be produced at least partly in the same sites  (e. g., massive stars and supernovae,
\citet{Timmes95}, \citet{Limongi06}).\\
For this study, we followed basically the same procedure as for the \Al\ one. We analyzed the same dataset in the 
 1170-1176 and 1330-1336 keV bands, with the same templates to estimate the \Fe\ flux.
The background determination method (Section~\ref{sec_intensity}) leads to the identification of 2900 and 2961 time-segments respectively.
Note that the contribution of the \Al\ line through its interactions with 
the detectors (Compton effect, diffusion,...) has to be taken into account for the \Fe\ analysis.
Being a strong line, \Al\ photons can interact inside the camera through the Compton effect and diffusion (the instrument response is non-diagonal)
and produce a continuum from 20 keV to 1.8 MeV in the data-space. 
This emission component is thus convolved with the instrument response to predict the corresponding counts in the data-space, and the predicted counts in the 1170-1176 and the 1330-1336 keV bands are included in the sky model during the \Fe\ fluxes analysis.
 Moreover, the  contribution 
from the diffuse continuum has been taken into account assuming the power law model
determined in a previous work \citep{Bouchet11}.
Note that these effects are very small, well below the statistical error bars.
\\
For all the tested distributions, the \Fe\ isotope lines are detected at a level of 
2$\sigma$ and 3$\sigma$ respectively, in the 1170-1176 and 1330-1336 keV bands 
(Table~\ref{table:radiocativesbis}). Their global mean-flux in the  inner Galaxy is about 
4\dixmoinscinq. 
Systematics in the flux determination of these large-scale structures due to the background pattern determination method is below  $\sim$25$\%$
for the \Al\ and around $\sim$30$\%$ for the \Fe\ lines (fluxes obtained with the fixed-pattern  are systematically higher). 
\\
As a conclusion, even though poorly constrained, the \Fe\ to \Al\ ratio obtained during this analysis is  
around 0.14 , which agrees with values previously obtained by \citet{Smith04} 
and \citet{Wang07}.  
%
\section{Discussion and Conclusion}\label{sec4}
%
For more than a decade, the reference map for the spatial morphology of the \Al\  
emission was provided by  \COMPTEL.  
Several maps have been published, along the CGRO mission, 
with more and more data, but also different
analysis methods.
Consequently, while 
the extended morphology of the emission is assessed and results compatible, some local features not always appearing, depending on the method. 
Indeed, the first MEM images \citep{Diehl95, Oberlack97} exhibit  many low-intensity structures. 
A small number of them remain in the latest MEM image \citep{Pluschke01}, giving some confidence 
in their reliability.  In parallel, the chief 
features of these MEM images were confirmed with a MREM  algorithm \citep{Knodlseder99b}.
This algorithm is based on an iterative expectation maximization scheme and a wavelet filtering algorithm. 
This wavelet filter suppresses the low-significance features, which are potential artifacts, by applying a user-adjustable threshold. This aims to produce the smoothest image consistent with the data \citep{Knodlseder99b}.
Note that the early MREM map built from \SPI\ data by \citet{Knodlseder06}
 was too rough to bring any additional information compared to  \COMPTEL\ ones. \\
Now that more than 10 years of \SPI\ data are available, 
our main goal was to refine   
the \COMPTEL\ view of the \Al\ line emission at 1.8 MeV  and to investigate the related \Fe\ lines around 1.2 and 1.3 MeV. 
To study their spatial morphology,  we have developed specific tools.
We have performed two kinds of image reconstructions. The first one is based on existing maps,
 at various wavelengths, 
which are used as a sky model, convolved with the instrument response and compared to the data (model-fitting). 
The second one consists in direct sky reconstruction from the data, implying an inverse-problem. 
While the first method reveals the global large scale morphology  of the emission, the latter 
allows us to look for small scale structures like local regions of \Al\ production.\\
The \Al\ line is detected at \(\sim 20 \sigma\), to compare to the \(\sim 30 \sigma\) obtained with 
the analysis of \COMPTEL\ data obtained in 5 years \citep{Knodlseder99}.\\
In addition, comparing  the  MREM \COMPTEL\ image \citep{Knodlseder99b} and \SPI\ one, we report that a number of structures appear in both analyzes.
Note that the \SPI\ sky exposition is non-uniform, mostly concentrated along the Galactic plane and  differs from the \COMPTEL\ one. Thus, \SPI\ is more sensitive in the Galactic plane than at high-latitudes, with a
point-source (3 \(\sigma\)) sensitivity  of 1.4\dixmoinscinq\ in the Galactic Center region to be compared to the 0.8 to 1.4\dixmoinscinq\ reported 
by \citet{Pluschke01} for \COMPTEL.
\subsection{Data analysis : issues and solutions}
A first point to mention is that, for \SPI\ as for \COMPTEL, the background treatment is a 
tricky issue. 
In \SPI\ data, to disentangle the signal and the instrumental background contributions, we rely on the 
ability of the instrument to measure simultaneously both of them, due to the properties of the coded-mask aperture imaging system. 
Moreover, the evolution of the background intensity with time can be determined with a segmentation 
code developed specifically to take into account the background
 variation. The major advantage is that it strongly reduces the 
number of parameters  related to the background. 
In a second step, we
determine the background detector pattern by assuming that high-latitude exposures 
constitute a good approximation of ``empty-fields''. 
However, they  may contain signal from the diffuse  emission (Section~\ref{sec_uniformity})
We have thus implemented the possibility  to determine the background pattern during the data-reduction process, 
assuming that the signal is sufficiently well known. 
The two approaches for background pattern determination have been compared to assess the robustness of the 
results, leading us to essentially the same conclusions regarding the \Al\ emission characteristics.\\
\subsection{\Al\ and \Fe\ large-scale emission}
In the model fitting  analysis based on other wavelength maps, our data do not allow us 
to distinguish a unique preferred template since several ones lead to similar likelihood parameter values. 
From \COMPTEL~data, \cite{Knodlseder99} had concluded that the best tracers were 
DIRBE 240$\mu$  and 53 GHz free-free maps. 
We agree that the FIR maps with wavelengths 60 to 240 $\mu$ 
or 53 GHz (free-free and dust) appear  statistically as the best estimates of the \Al\ emission global morphology 
but  CO, NIR A[4.9$\mu$] and A[3.5$\mu$]  extinction-corrected maps have also to be considered  with only 
a slightly lesser degree of confidence. In addition, ``low-contrast'' maps  observed at 25$\mu$ 
and  12$\mu$  provide an equally good description of the emission.\\
Indeed,  our results  confirm  that the \Al\ emission follows more or less 
  the distribution of the extreme Population \(I\), the most 
massive stars in the Galaxy \citep{Diehl95}. It is known that the massive stars, 
supernovae and novae produce the long-lived isotopes \Al\ and \Fe\ with half-lives of 0.7 
and 2.6 My. 
On the another hand, large amounts of dust/grains condense in core collapse SNe ejecta while 
massive star-supernovae are  major dust factories, therefore dust FIR and MIR maps are expected to
correlate with the corresponding emissions.
\\
Quantitatively, the flux extracted through the model fitting method for the inner Galaxy 
($\vert l \vert \leq 30^{\circ}$,$\vert b \vert \leq 10^{\circ}$)
is  found to be around  3.3$\times$ 10$^{-4}$\phf 
while it does not depend much on the  sky model (particularly if we restrict ourselves to the preferred ones).
 This flux is consistent with earlier measurements of both \INTEGRALSPI\ and 
\COMPTEL~instruments
 and corresponds to a total \Al\ mass contained in our Galaxy of \(\sim\)3$M\sun$,(but,  see also
\citet{Diehl10} and \citet{Martin09}).
However, \citet{Churazov10} using a  ''light-bucket'' method obtain a flux  of 4 $\times$ 10$^ {-4} $ 
\phf for the region delimited by $\vert l \vert \leq 40^{\circ}$,$\vert b \vert \leq 
40^{\circ}$. For the same region, we obtain a flux of $\sim$ 3.9 $\times$ 10$^{-4}$\phf with the most 
probable spatial morphologies.\\
Another observable quantity, related to \Al\ production, is the ratio between  \Fe\ and \Al\ line fluxes in the inner Galaxy. 
 From a theoretical point of view, the 
\Fe\ to \Al\ flux ratio provides a test for stellar models, as predictions of the yields 
of massive stars depend strongly on the prescription of nuclear rates, stellar winds, 
mixing and rotation \citep{Woosley07,Tur10}. 
With our analysis, the \Fe\ mean flux in the inner Galaxy is 
about 4\dixmoinscinq, leading to a  \Fe\ to \Al\ ratio between 0.12 and 0.15. 
 Considering the  uncertainties still 
affecting the models, this results confirms those reported by \citet{Harris05, Wang07}, and can
be used to reject some 
hypotheses, but not yet to definitively discriminate the good ones. 
\subsection{Imaging results}
To go further in the details of the emission distribution, the flux extraction must be independent  of
any template, and rely on a direct imaging reconstruction of the \Al\ emission.
To realize it, we chose the MEM tool because of its ability to process high-resolution images. 
Our MEM code is based on \citet{SB84} algorithm. The main improvement we implemented to the basic algorithm 
is the possibility to refine the background determination during the iterative computation of
 the solution.\\
The resulting \SPI\ image presented in Figure~\ref{fig:image_alu} resembles  the  MEM \COMPTEL\ images in 
term of angular resolution and details. With a resolution  fixed to 
 $\sim$6$^\circ$, it gives essentially the same information as 
the previous method, i.e. the emission is mainly confined in the inner-Galaxy disk (with $\vert b \vert \lesssim 7 ^\circ$).
But, it also 
suggests the presence of extended (a few degrees) emitting areas 
and makes it  quite instructive  to  compare the main  features with those observed in \COMPTEL\ maps: 
Several excesses appear in both \SPI\ and \COMPTEL\ data analyzes of \citet{Pluschke01},
 which reinforces their reliability see (Table~\ref{table:emission_regions}).\\
In the Cygnus region, Cyg OB2 cluster detection has been reported in \COMPTEL\ and \SPI\ data \citep{Martin09}.
We find at the same position a flux of $4.1-4.5 \pm 1.5$\dixmoinscinq\ compatible with the value of $3.9 \pm 1.1$\dixmoinscinq\ reported by \citet{Martin09}
and $3.7 \pm 1.1$)\dixmoinscinq\ reported by \citet{Pluschke00}.
However, a more complete analysis based on the excesses visible in the image reveals several spots in the Cygnus region and around (Table 2), suggesting a
complex structure of this area.\\
The Vela and Carina regions are detected, in our work, at  a low significance level (2$\sigma$).
For Vela, the measured flux is comparable to the value reported by \citet{Diehl95b}. For Carina, the flux
is comparable to the estimation of \citet{Voss12} using \SPI\ data and  \citet{Knodlseder96} using \COMPTEL\ data.\\
In the Sco-Cen region, emission, not detected in the \COMPTEL\ data, was reported in the
\SPI\ data at a high significance level ($6 \pm 1$\dixmoinscinq) by \citet{Diehl10} with a somewhat different analysis. Using 
their location and spatial extension, we find a flux of $4.2 \pm 1.6$\dixmoinscinq\, in the most favorable case  
(the flat H$_I$ map used to model the diffuse large-scale emission).  In our image, the excess appears shifted
and less extended (Table 2). Moreover, the measured flux depends strongly on the  template used to model the large-scale emission of the \Al\ line.
This lead us to consider that this result requires additional work to be confirmed.\\
On the other hand, 
we pick up significant and robust excesses, not explicitly reported by the \COMPTEL\ team.
Two of them, rather extended, are seen in the Taurus/Anticenter region at (l,b)$\simeq$ (161$^\circ$,-3$^\circ$) and (l,b)$\simeq$ (149$^\circ$, 8$^\circ$)
 with fluxes of $4-6 \pm 2$\ and $8-9 \pm 3$\dixmoinscinq. It is not excluded that they belong to a same structure, leading to a $\sim$4$\sigma$ detection. 
Furthermore, they can be correlated with a weak feature around (l, b) $\simeq$(160$^\circ$,0$^\circ$) in the \COMPTEL\ MREM image \citep{Knodlseder99b}.
A third feature appears at high-latitude (($7 \pm 2$)\dixmoinscinq).
Obviously, we can not exclude that this spot is due to statistical data noise,
but its stability against the various analysis methods supports its reliability.\\
We have to point out that
the present analysis has been optimized for 
large scale emissions (handling the all-sky data set) while any "local" analysis requires a dedicated procedure. 
 Briefly, in this case, we have to use the exposures whose pointing 
 direction are not too far from the source or region of interest 
 (less than $12^{\circ}$) and to refine the sky model,  to optimize the signal-to-noise ratio.\\
\citet{Chen96} (see also \citet{PD95}) have proposed an interpretation of the \COMPTEL\ \Al\ emission by linking 
enhanced emission spots to the Galactic spiral arms. 
This relation has been confirmed with the  observation of the line shift along the Galactic plane in the \SPI\ data \citep{Diehl06, Kretschmer13}.
Some of the identified regions match excesses we point out  in our 
analysis, reinforcing the reliability of the emissions and supporting the proposed scenario (labeled with ' * '  in  Table 2). \\
To conclude, the \SPI\ instrument gives us  one of the rare opportunities to get  pieces of information about 
the spatial distribution of \Al\ line and the nucleosynthesis process. Fifteen years after the first results obtained by \COMPTEL, 
we confirm the chief features as well as some of  the low-intensity structures reported by this instrument
through direct-imaging of the sky and templates  comparison. 
While we cannot hope to increase significantly the amount of data recorded by \INTEGRAL\ (more than 10 years 
of  observation have been included in the presented analysis), some improvements in the data analysis appear achievable
 to still refine our results. 
In particular, an accurate  modeling of the response outside the field of view is a prime objective  
for studying emissions above  $\sim$ 1 MeV together with the analysis of the multiple events, which contain
$\sim$20$\%$ of the $\gamma$-ray photons.

\begin{acknowledgments}
We thank the anonymous referee for suggestions and constructive 
comments.
We would like to thank James Rodi for careful reading  of the manuscript.
The \INTEGRALSPI\ project has been completed under the responsibility and 
leadership of CNES. 
We are grateful to ASI, CEA, CNES, DLR, ESA, INTA, NASA and OSTC for support.
\end{acknowledgments}

\newpage
\appendix
\section{The Maximum Entropy algorithm}
\label{appendixA}
The Maximum-Entropy algorithm aims to maximize the following function Q (Equation 4):
\begin{equation}
Q(f)= \alpha H(f,m)-\frac{C(f)}{2} 
\label{eq:A1}
\end{equation}
where \(\alpha\) is a regularization parameter. 
For an image containing \(N \) pixels, the following entropy function \(H\) is used 
\begin{equation}
 H(f,m)=\sum_{i=1}^{N} (f_i-m_i-f_i log \frac{f_i}{m_i})
\label{eq:A2}
\end{equation}
where \(m_i\) is a model image assigned to pixel \(i\), which expresses our prior knowledge about the sky intensities \(f_i\).The sky intensities and their model are positive quantities. \(C(f)\) measures the discrepancy between the measured data d and the reconstructed model of the data. A single statistical constraint is generally used and for data with a Gaussian noise, C is the chi-square function:
\begin{equation}
C=\sum_{i=1}^{M} \frac{(\sum_{j=1}^{N} R_{ij} f_j -b_i-n_i)^2}{\sigma_i^2}
\label{eq:A3}
\end{equation}
In this expression \(M\) is the number of measured data-points \(n_i\), \(\sigma_i^2\) represents the noise level and \(b_i\) the instrumental background for each data-point. \(R\) is a matrix of size \(M \times N\), which represents the response of the instrument. To simplify the presentation of the algorithm, the values \(b_i\) are supposed to be known and provided by the user.\\
The resulting map \(f\) obtained by maximizing Equation~\ref{eq:A1} will be unique since the surfaces \(H=constant\) and \(C=constant\) are both convex. For every fixed constraint level, \(C_{aim} \), 
\begin{equation}
C(f) = C_{aim}
\label{eq:A4}
\end{equation}
defines an ellipsoidal hyper-surface of radius \(C_{aim}\) in the N-dimensional image-space. For this surface there is only one tangent point \(f\) with a certain (maximal for Equation~\ref{eq:A4}) entropy hyper-surface for which the gradient $\nabla$Q vanishes. Therefore, f is the maximum entropy solution for Equation~\ref{eq:A4} given by the formula:
\begin{equation}
\nabla Q=\alpha \nabla H -\frac{1}{2} \nabla C=0
\label{eq:A5}
\end{equation}
%
\subsection{Convergence method}
An iterative procedure is needed to solve the set of non-linear equations (Equation~\ref{eq:A5}).
It can be the following: Starting from the point of absolute maximum-entropy solution ( \(f^0=m\), \( \alpha=\infty \) ), one finds at each iteration, the maximum entropy point \( f \) , which fulfills the condition of Equation~\ref{eq:A5} for certain ellipsoidal hyper-surfaces (Equation~\ref{eq:A4}) with the condition:
\begin{equation}
C(f+\delta f) < C(f)
\label{eq:A6}
\end{equation}
The new solution is updated by incrementing the previous solution with a vector \( \delta f\). The solution is forced to follow the maximum-entropy trajectory by using the appropriate value of \(\alpha\) defined by Equation~\ref{eq:A5}.\\
As the solution is updated at each iteration, the approximation of \(Q\), through its quadratic expansion, should remain accurate. Maximizing \(Q(f)\) subject to \(|\delta f|^2 \leq l_0^2 \) for some small value of \(l_0\), between successive iterations, fulfills the requirement. In addition, the tendency of the search-direction algorithm to produce negative values of \(f\) is limited. However such distance limit tends to slow the attainment of high-values of \(f\).
To overcome this defect, \citep{SB84} (hereafter SB84) suggest to use, instead of the squared length of the increment:
\begin{equation}
l^2=\sum_i (\delta f_i / f_i)^2 \leq l_0^2
\label{eq:A7}
\end{equation}
Using a distance in this form is equivalent to put a metric with \(-\nabla \nabla H \) onto the image-space.
\begin{equation}
g_{ij}=1/f^i (i=j) \mathrm{~and~} g_{ij}=0 (i \neq j) 
\label{eq:A8}
\end{equation}
The upper index denotes a contra-variant vector.\\ However, the problem is a N-dimensional problem and can be difficult to solve for a large number of unknowns \(N\) with a Newton-Raphson procedure.\\
It is one of the reasons why we use the algorithm proposed by SB84. It consists in choosing, at each iteration, three search-directions \(e_1 \), \(e_2 \) and \(e_3 \) derived from the gradients \(\nabla C \) and \(\nabla H \), as well as from the matrix of curvature of C, \(\nabla \nabla C\). In this way, the N-dimensional problem is reduced to a three-dimensional problem. The three search-directions, by using the metric given by Equation~\ref{eq:A8}, are
\begin{equation}
e_1=f (\nabla H),~e_2=f (\nabla C),~e_3= |\nabla H|^{-1} f (\nabla \nabla C) f (\nabla H)-|\nabla C|^{-1} f (\nabla \nabla C) f (\nabla C)
\label{eq:A9}
\end{equation}
The entropy metric is used to define the lengths
\begin{equation*}
| \nabla S |=[\sum f^i (\frac{\partial S}{\partial f^i})^2]^{1/2},~|\nabla C|=[\sum f^i (\frac{\partial C}{\partial f^i})^2]^{1/2} 
\end{equation*}
The solution is updated by adding the increment vector as follows
\begin{equation}
f=f+\delta f= f+\sum_{\mu=1}^{3} x^{\mu} e_{\mu}
\label{eq:A11}
\end{equation}
The values to be found are the lengths \(x_{\mu}\) ( \(\mu\)=1,2,3) along the search directions \(e_{\mu}\). For this purpose, \(H\) and \(C\) are modeled by their Taylor 2 -order expansions \(\tilde H\) and \(\tilde C\) at the current solution \(f\)
\begin{equation*}
\tilde H(x)=H_0+H_{\mu} x^{\mu}- 1/2 g_{\mu \nu} x^{\mu} x^{\nu},~\tilde C(x)=C_0+C_{\mu} x^{\mu}- 1/2 M_{\mu \nu} x^{\mu} x^{\nu};~l^2=g_{\mu \nu} x^{\mu} x^{\nu}
\end{equation*}
where
\begin{equation*}
H_{\mu}=e_{\mu}^T \cdot \nabla{H},~C_{\mu}cC_{\mu}^T \cdot \nabla{C},~g_{\mu \nu}=e_{\mu}^T \cdot e_{\nu},
~M_{\mu \nu}=e_{\mu}^T \cdot \nabla \nabla C \cdot e_{\nu}
\end{equation*}
After simultaneous diagonalisation of the quadratic models of H and C, in a common basis, formed by the eigenvectors of the matrix \(g_{\mu \nu}\) computed on the basis formed by the eigenvectors of the matrix \(M_{\mu \nu}\), one obtains:
\begin{equation*}
\tilde H(x)=H_0+H_{\mu} x_{\mu}- 1/2 x_{\mu} x_{\mu},~\tilde C(x)=C_0+C_{\mu} x_{\mu}- 1/2 \gamma_{(\mu)} x_{\mu} x_{\mu}; l^2=x_{\mu} x_{\mu}
\end{equation*}
where \(\gamma_{(\mu)}\) are the eigenvalues of \(M_{\mu \nu}\). This is a low-dimensional problem and standard algorithms can perform these operations.
According to the maximum-entropy trajectory (\(\nabla Q=0\) as defined in Equation~\ref{eq:A5}), the step-lengths are 
\begin{equation}
x_{\mu}=\frac{(\alpha S_{\mu}-C_{\mu})}{(\gamma_{(\mu)}+\alpha)}
\label{eq:A15}
\end{equation}
SB84 introduce explicitly the distance constraint into the optimization process.The quadratic model of \(Q\) becomes
\begin{equation}
\tilde Q=\alpha \tilde H -\tilde C/2  + p l^2/2 \mathrm{~with~} p > 0
\label{eq:A16}
\end{equation}
Then, the step-lengths \( x_{\mu} \) are  given by
\begin{equation}
x_{\mu}= \frac {\alpha S_{\mu}-C_{\mu}}  {p+\gamma_{(\mu)}+\alpha}
\label{eq:A17}
\end{equation}
In SB84, \(p\) can be interpreted as an artificial increase of each eigenvalue \( \gamma_{(\mu)}\) of C. The equivalent quadratic model of \(C\) is 
\begin{equation}
C_p(x)=C_0+ C_{\mu} x_{\mu} +1/2 (\gamma_{(\mu)}+p) x_{\mu} x_{\mu}
\label{eq:A18}
\end{equation}
%
\subsection{Control of the algorithm}
%
The control of the convergence of the algorithm is performed directly on the constraint C, and not on the Lagrange multiplier \( \alpha \). The aim is to maximize \( S \) over \( C=C_ {aim} \). The minimum reachable value of \(C\) is
\begin{equation*}
C_{min}=C_0-\frac{1}{2} \sum_{i=1}^3 \frac{C_{\mu}^2}{\gamma_{(\mu)}+p}
\end{equation*}
SB84 recommend to have a more modest aim \( \tilde C_{aim}\), for example
\begin{equation*}
\tilde C_{aim}=max( 2/3 C_{min}+ 1/3 C_0, C_{aim})
\end{equation*}
\(\tilde C(x)\) is an increasing function of \(\alpha\). The simultaneous control of \(\tilde C(x)\) and \(l^2\) is done through the value of \(p\), \(\tilde C(x)\) is an increasing function and \(l^2\) a decreasing function of \(p\).
By adjusting the values of \(\alpha\) and \(p\), one can reach the aimed result. The procedure is detailed in SB84.
%
\subsection{Stopping the algorithm}
In the '' historical'' version of the maximum-entropy, this process is repeated until \(C\) reaches the stopping value \(M\). SB84 suggest to reach at least a value lower than the largest acceptable value at 99 \% confidence,  about \( M+3.39 \sqrt M\) where \(M\) is the number of observations.\\
 Equation~\ref{eq:A5} forces the  gradients $\nabla$H and $\nabla$C  to be parallel.  Then, the parameter \(\alpha\) can be interpreted as the ratio of their lengths.
In addition, the algorithm is always checked by measuring the degree of non-parallelism between \(\nabla H\) and \(\nabla C\) through the value of
\begin{equation}
TEST=\frac{1}{2} | \frac{\nabla H}{|\nabla H|}-\frac{\nabla C}{|\nabla C|}|^2 
\label{eq:A21}
\end{equation}
The value is zero for a true maximum entropy image. SB84 indicate that reaching \( TEST \lesssim 0.1\) or so at the correct value of C, demonstrates that the unique maximum-entropy reconstruction has been attained.\\
In our application, we always reached \(C \le M \) and \(TEST \le 0.1 \). The function \(C\) always decreases  when the  iteration number increases. From a number of iterations, the value of TEST starts to decrease in a monotonic way and reaches a minimum value. From this point, its value becomes almost stable and in the worst case increases slightly whereas there is essentially no progress in the decrease of the value of \(C \).
In addition, we compare the function C to its quadratic approximation to ensure its validity. To achieve and maintain this constraint during the iterations, the length of the increment (Equation~\ref{eq:A7}) is kept small enough, this results in a larger iteration number. In general, 30 to 50 iterations are required (less if the increment constraint is higher or relaxed at any given number of iterations, but at the expense of an inaccurate 2nd-order approximation of C).
%
\subsection{Our application}
Different choices of C (f) are possible \citep{Bryan80}. We used a modified version of the \(\chi^2\) statistic, accurate for low numbers of counts, following the prescription of \citet{Mighell99}.  The image model is computed at each iteration as 
\begin{equation*}
m_i= cste= exp ( \frac { \sum_{i=1}^N f_i ln f_i} {\sum_{i=1}^N f_i})
\end{equation*}
\section{Simulations of imaging reconstruction}
\label{appendixB}

\begin{figure*}
\centering
\subfigure[Image template (shown at 6$^{\circ}$ FWHM)]{\includegraphics[width=\textwidth]{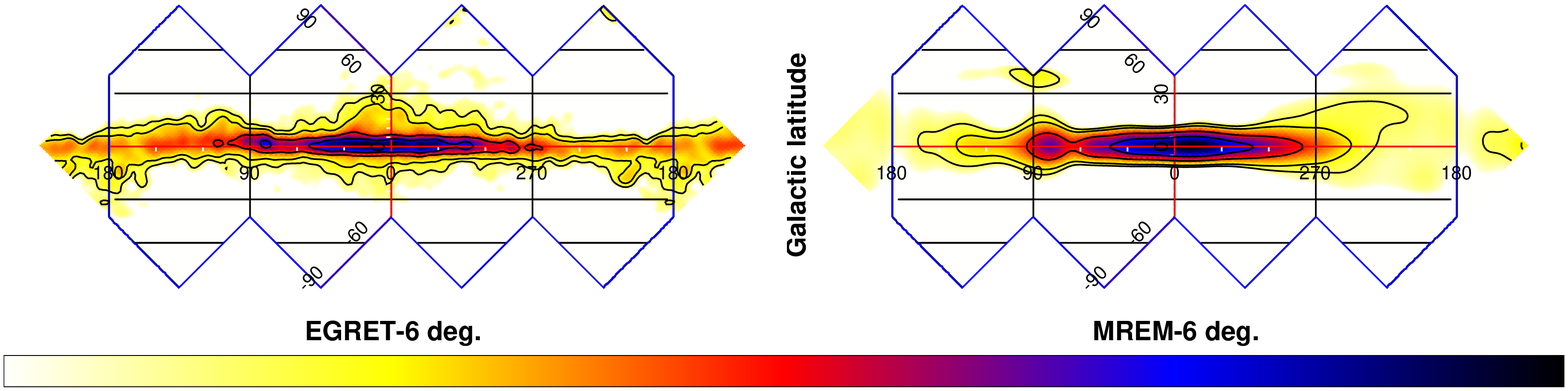}}
\hfill
\subfigure[Reconstruction 6$^{\circ}$ FWHM]{\includegraphics[width=\textwidth]{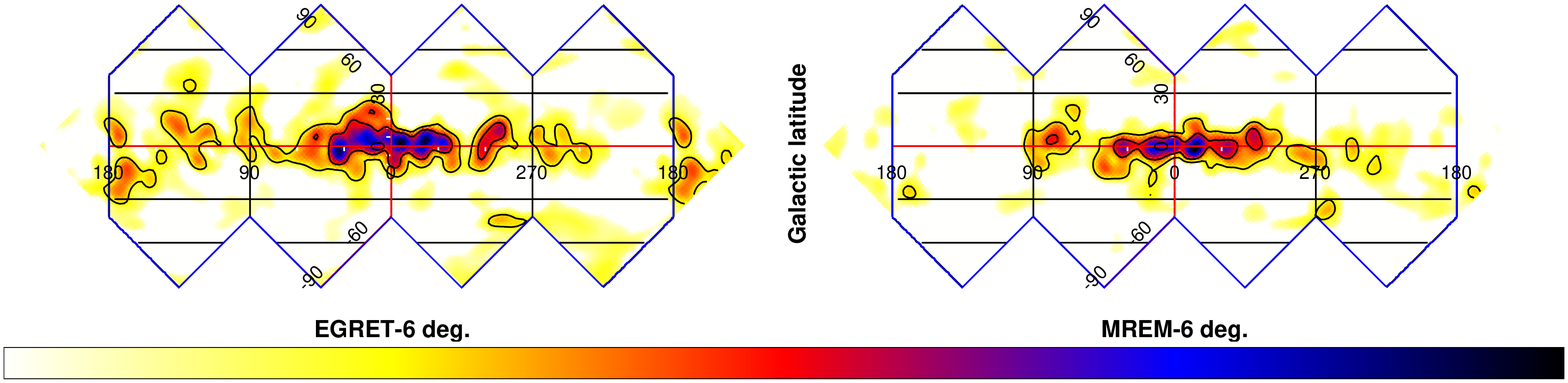} }
\hfill
\subfigure[Reconstruction 3$^{\circ}$ FWHM]{\includegraphics[width=\textwidth]{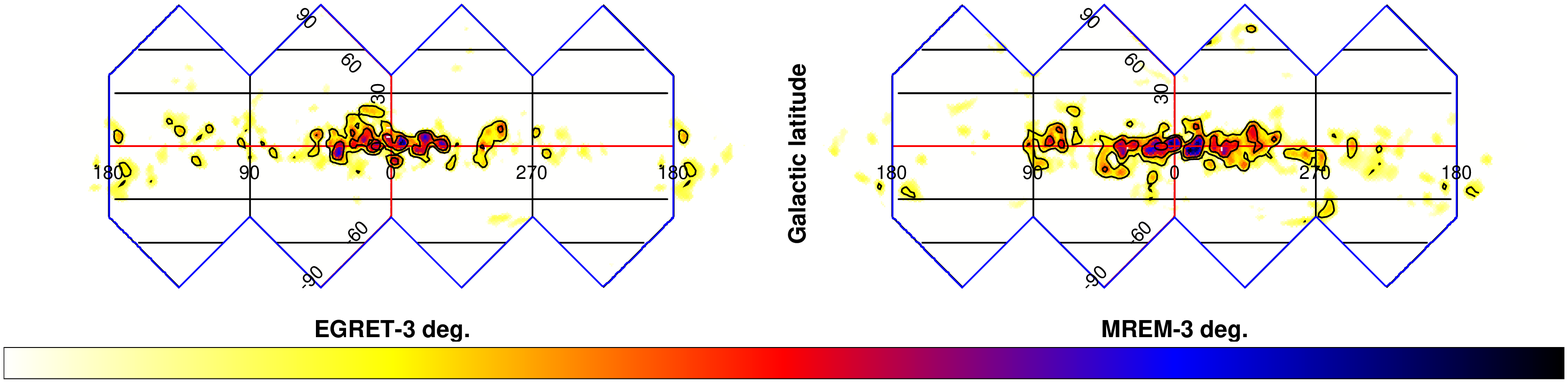} }
\caption{Simulated and reconstructed maps. The image contains 49152 pixels (pixel size is 0.9$^{\circ}$) ordered following the Healpix scheme. 
(Top) Simulated MREM (initially 3.8$^{\circ}$) and  EGRET (initially 2$^{\circ}$) displayed using
an angular resolution of 6$^{\circ}$ FWHM.  (Middle) The reconstructed  MREM  and EGRET fixing the output map resolution to 6$^{\circ}$ FWHM and (Bottom)
to 3$^{\circ}$  FWHM. Images are displayed using the DS9 software (http://ds9.si.edu/site/Download.html).
The images are scaled between their maximum intensity \(f_{max}\) and  \(f_{max}/25\). The scale is in square root unit of the flux 
(ds9 Scale Square Root menu).
(Top) contours correspond to \(f_{max}/2\), \(f_{max}/4\), \(f_{max}/10\) and \(f_{max}/16\). 
(Middle) \(f_{max}/2\), \(f_{max}/4\) and \(f_{max}/10\).
(Bottom) \(f_{max}/5\) and \(f_{max}/12\).
}
\label{fig:appendixB}
\end{figure*}

\begin{figure*}
\centering
\subfigure[EGRET simulated template]{\includegraphics[width=0.49\textwidth]{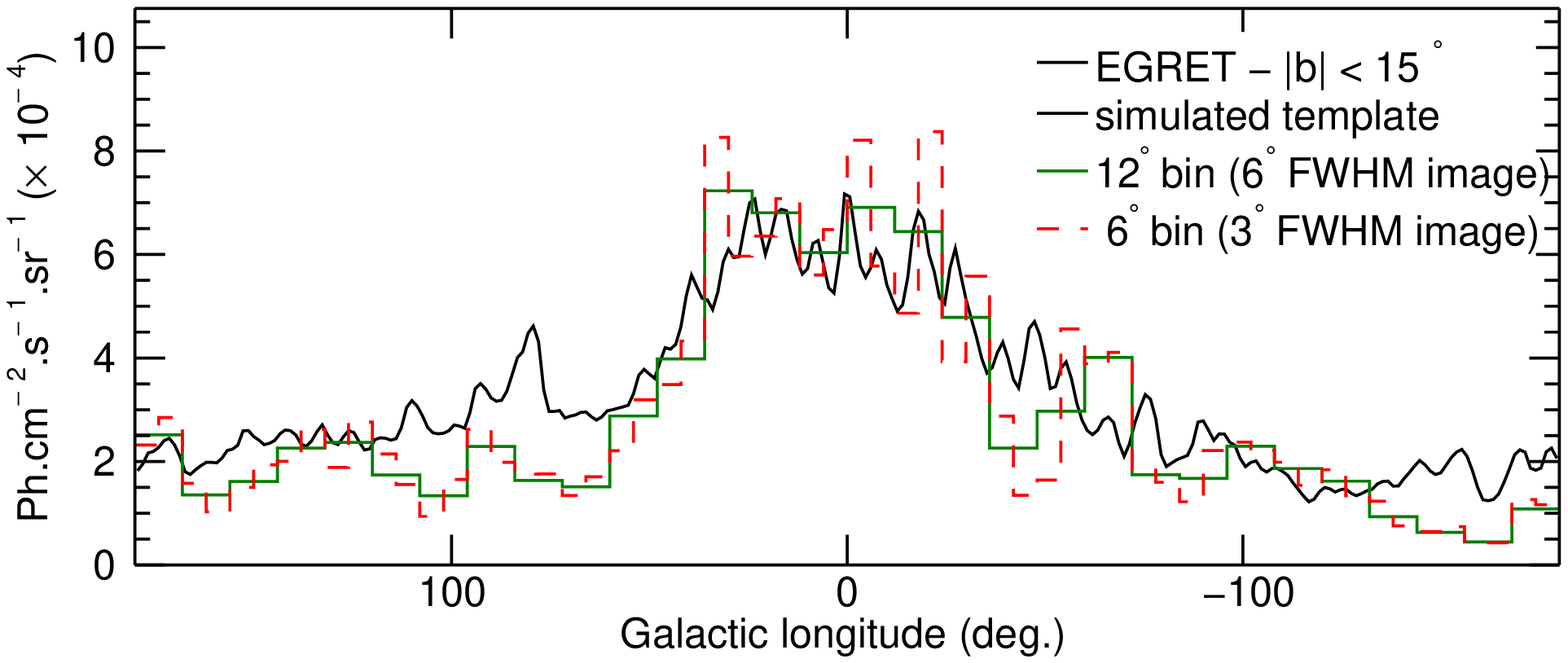} }
\hfill
\subfigure[MREM simulated template]{\includegraphics[width=0.49\textwidth]{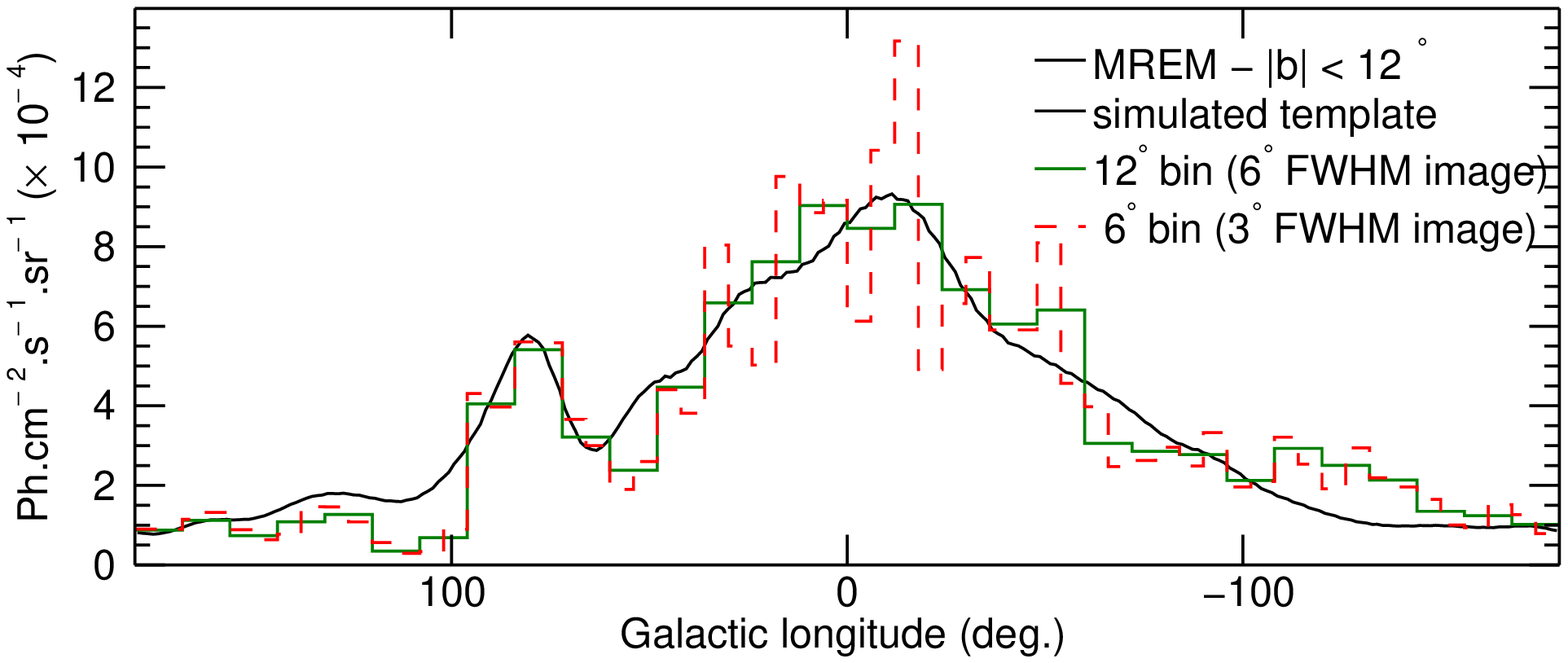}}
\caption{Longitude profiles of the reconstructed images compared to the simulated images (See Figure~\ref{fig:appendixB}.}
\label{fig:lprofiles}
\end{figure*}

To simulate synthetic data, we rely on the model-fitting analysis. A given template map is used to model the distribution of the \Al\ line over the Galaxy (Section 3.2). Expected counts are obtained by convolving this sky model with the instrument response and then adding the background model. The intensity of the map and the intensity variations of the background are the parameters which are adjusted to the recorded data.\\
 Our simulations start from the predicted counts.
Poisson statistical fluctuations are added to them to build the simulated data, which are in turn analyzed similarly as the real data. We  simulated data from the MREM template, which represents a smooth input map, and EGRET template, which is a more structured. The input maps and their reconstruction are displayed 
on Figure~\ref{fig:appendixB}.
We present two  image reconstructions which differ in the resolution of the final image, forced to the values  of 3$^{\circ}$ and 6$^{\circ}$ (FWHM).
In all cases, the fluxes simulated and measured in the inner Galaxy are recovered within 5$\%$ for both MREM and EGRET simulated maps. 
From these simulations, we conclude that statistical noise produces structures whose intensity can reach about one-tenth of the image maximum intensity. 
Above this value, there are no particular structures  created during the image reconstruction process at any particular position. 
The longitude profiles of the reconstructed images are compared to the simulated images on Figure~\ref{fig:lprofiles}.

\newpage
\clearpage
%
%
%
%
%
\begin{deluxetable}{l c  c c}
\tablecaption{Maps used as templates.}
\tabletypesize{\footnotesize}
\tablehead{
&\colhead{  Tracer }
&\colhead{  Mechanism }
}
\startdata
$\dag$MIR 12 and 25$\mu$            & warm dust (T\(\sim\)250 and \(\sim\)120 
K) & dust nano-grains and PAHS         \\
                                     &             /AGBs                     & 
heated to high temperature        \\
$\dag$53 GHz sync                    & cosmic-rays/magnetic field             & 
Synchroton                       \\
$^a$\COMPTEL-MEM                      &                                        & 
                                 \\
IC                                   & Inverse-Compton from GeV               & 
Inverse-Compton                  \\
                                     & electrons on the CMB and ISRF          &  
                                \\
$\dag$NIR 1.25,           4.9 $\mu$  & stars (K and M giants)                 & 
star light                       \\
$^b$H$_I$ (21 cm)                    & H hyperfine transition                  & 
Neutral hydrogen                 \\
$\dag$NIR   2.2$\mu$                 & stars (K and M giants)                 & 
star light                       \\
$\dag$EGRET                          &  Interstellar gas/cosmic-rays          & 
nuclear interactions             \\
$\dag$NIR 3.5   $\mu$                & stars (K and M giants)                 & 
star light                       \\
$\dag$53 GHz free-free               & ionized gas                            & 
Free-free                        \\
$\dag$53 GHz dust                    &  dust                                  & 
Thermal dust                     \\
$\dag$FIR 100, 140 and 240$\mu$  & warm dust (T  \(\sim\)30,                
& microns sized dust emitting in thermal \\
                                     &           \(\sim\)21 and \(\sim\)12 K)    
 & equilibrium with the heating ISRF       \\     
$^a$\COMPTEL-MREM                     &                                        & 
                                \\
$\dag$FIR 60$\mu$                     & warm dust (T\(\sim\)50 K)                
& microns sized dust emitting in thermal \\
                                     &                                        & 
equilibrium with the heating ISRF       \\     
$^c$CO                               & CO rotational transition               & 
Molecular gas / young stars                   \\
$^d$NIR extinction-corrected map     &  stars (K and M giants)                & 
star light                        \\
3.5, 4.9 $\mu$ (hereafter A[${3.5\mu}$ and \Ared4.9 )  &                      &  
                                 \\
\enddata
\tablecomments{
The maps are ordered in ascending ``contrast'', defined here as the 
ratio between the fraction of the emission enclosed in the region $\vert l \vert 
< 150^{\circ}$, $\vert b \vert < 
15^{\circ}$  to that of the whole sky. The value of the ratio varies from 0.4 (25$\mu$) 
to nearly 1 (\Ared4.9).\\
$\dag$ available at http://lambda.gsfc.nasa.gov or http://heasarc.gsfc.nasa.gov.\\
The NIR, MIR and FIR maps are the COBE/DIRBE Zodi-Subtracted Mission Average (ZSMA) maps, from which
zodiacal light contribution has been subtracted. Our 53 Ghz synchrotron  presents some inaccuracies and is used 
for indicative purpose.
 $^a$ The maximum-entropy (MEM)  and  
Multiresolution Regularized Expectation Maximization (MREM)
all-sky image of the Galactic 1809 keV line 
emission observed with \COMPTEL\ over 9 years 
\citep[][and references therein]{Pluschke01}.$^b$\citet{Dickey90}. 
$^c$\citet{Dame01} with central peak removed (Section~\ref{sec_templates}).
$^d$The NIR extinction maps (Section~\ref{sec_templates}).
}
\label{table:tracers}
\end{deluxetable}
%
\begin{deluxetable}{l c c c c}
\tablecaption{\(^{26}\)Al emission sites and flux in units \dixmoinscinq.}
\tabletypesize{\scriptsize}
\tablehead{ 
\colhead{ Source }
&\colhead{ Position }
&\colhead{ Spatial  }
& \multicolumn{2}{c}{\Al\ large-scale morphology} \\
\colhead{ name }
&\colhead{  l$^{\circ}$, b$^{\circ}$  }
&\colhead{ morphology}
&\colhead{ IC map }
&\colhead{  \Ared4.9}
}
\startdata
\multicolumn{5}{c}{Cygnus region} \\
Cyg OB2 cluster$^d$  & \textbf{81,-1} & \textbf{Gaussian, \(\sigma\)=3} &   4.5 $\pm$ 1.5  & 4.1 $\pm$ 1.5  \\
\multicolumn{5}{c}{Mutiple sources - most significant excesses} \\
         & $^\dag$81,-1    &  Gaussian, \(\sigma\)=3               & 4.3 $\pm$ 1.6    & 3.8 $\pm$ 1.6    \\     
                & 100, 6   &  Point-source                         & 2.9 $\pm$ 1.1    & 2.7 $\pm$ 1.1     \\
         & $^\dag$64, 1    &  Point-source                         & 2.2 $\pm$ 1.1    & 2.0 $\pm$ 1.1     \\
%
\multicolumn{5}{c}{} \\
Sco-Cen$^{a,+}$  & \textbf{350, 20}  & \textbf{disk r=10$^{\circ}$} & 0.4 $\pm$ 1.4 & 1.9 $\pm$ 1.4   \\
                 & 360, 16  &  disk r=5$^{\circ}$                   & 0.5 $\pm$ 0.9 & 1.1 $\pm$ 0.9   \\

\multicolumn{5}{c}{Orion/Eridanus region} \\
Orion/Eridanus$^e$ &  & \textbf{150 $<$ l $<$ 210, -30 $<$ b $<$ 5}    &   1.2 $\pm$ 3.2  & 0.5 $\pm$ 3.1   \\
\multicolumn{5}{c}{} \\
Carina$^b$    & $^*$\textbf{287,0}  & \textbf{disk r=3$^{\circ}$} &   3.1 $\pm$ 1.5  & 2.8 $\pm$ 1.5    \\
Vela $^c$     & $^*$\textbf{267,-1} & \textbf{disk r=4$^{\circ}$} &   3.6 $\pm$ 1.8  & 3.3 $\pm$ 1.8   \\
\multicolumn{5}{c}{} \\
Taurus$^\dag$  & $^*$161,-3   &  Disk r=5$^{\circ}$                &   5.6 $\pm$ 2.1  & 4.4 $\pm$ 2.1   \\
               & 149, 8   &  Disk r=11$^{\circ}$                   &   8.5 $\pm$ 2.9  & 7.9 $\pm$ 2.9   \\    

\multicolumn{5}{c}{} \\
\multicolumn{5}{c}{Other significant excesses} \\
               & 226, 76  &  Gaussian, \(\sigma\)=3$^{\circ}$     & 7.2 $\pm$ 1.8    & 6.8 $\pm$ 1.8     \\ 
%
\multicolumn{5}{c}{Known high-energy emitting sources} \\
Crab           & \textbf{185,-6}    &  \textbf{Point-source}      &   1.2 $\pm$ 1.3  & 1.1 $\pm$ 1.3      \\
Cyg X-1        & \textbf{71, 3}     &  \textbf{Point-source}      &   0.2 $\pm$ 1.1  & 0.2 $\pm$ 1.1      \\
PSR B1509-58 & $^*$\textbf{320,-1.2} & \textbf{Point-source}      &   2.8 $\pm$ 0.9  & 2.8 $\pm$ 0.9      \\
\enddata
\tablecomments{
The first column contains the name of the source if it is known. The second column the position of the source in galactic coordinates and the third column the source spatial morphology; a point (point-source), an axi-symmetric Gaussian (\(\sigma\) indicated) or a disk (radius indicated)). The position and spatial morphology in bold correspond to  known sources or are based on published works.
 Next columns give the source fluxes obtained for two template maps used to model the large-scale distribution of \Al\ over the Galaxy; IC (low-contrast map) and  \Ared4.9\  (high-contrast map). 
A model, comprising the \Al\ line distribution, the sources and the background, is adjusted to the data.
The uncertainties are obtained from the model-fitting analysis by 
using the curvature matrix of the likelihood function and terms 
associated with the variance of the solution.
The different labels identify excesses, which have been already mentioned in the literature by: 
$^a$ \citep{Diehl10} - 
$^b$ \citep{Voss12} - 
$^c$ \citep{Diehl95b} - 
$^d$\citep{Martin09} - 
$^e$\citep{Diehl02}. \\
$^*$ Possible association with the spiral arms \citep{Chen96} - 
$^\dag$ Suspected or visible in \COMPTEL\ $^{26}$Al image \citep{Oberlack97, Pluschke01}.
$^+$See Section~\ref{region_excesses}.
}
\label{table:emission_regions}
\end{deluxetable}
%
\begin{deluxetable}{l c c c c c c c}
\tablecaption{Radioactive line fluxes in the inner Galaxy in units of $\times$10$^{-4}$ \phf.}
\tabletypesize{\scriptsize}
\tablehead{
&\colhead{  \(25\mu\) }
&\colhead{ \COMPTEL-Al }
&\colhead{ \(240\mu\) }
&\colhead{ \COMPTEL-MREM }
&\colhead{ \(60\mu\) }
& \colhead{ CO }
& \colhead{\(\dag\) \Ared4.9 }
}
\startdata
{ \Al\ }  & $ 3.68 \pm 0.21$ & $3.41 \pm 0.20$ & $3.15 \pm 0.18$ & $3.24 \pm 0.19$ & 
$3.11 \pm 0.18$ & $3.05 \pm 0.18$ & $ 2.92 \pm 0.17$  \\
{\Fe}     & $ 0.47 \pm 0.16$ & $0.45 \pm 0.15$ & $0.38 \pm 0.14$ & $0.38 \pm 0.14$ & 
$0.38 \pm 0.13$ & $0.41 \pm 0.13$ & $ 0.42 \pm 0.13$  \\                  
{ \Fe/\Al} & $0.13 \pm 0.05$ & $0.13 \pm 0.05$ & $0.12 \pm 0.05$ & $0.12 \pm 0.05$ & 
$0.12 \pm 0.05$ & $0.14 \pm 0.05$ & $ 0.15 \pm 0.06$   \\
\enddata
\tablecomments{
Fluxes are given for the inner galaxy ($\vert l \vert < 30^{\circ}$ and $\vert b 
\vert < 10^{\circ}$). $\dag$ the map is corrected from reddening and fluxes are 
essentially zero for  $\vert b \vert > 20^{\circ}$. \Fe\ is the average 
flux of the 1173 and 1333 keV lines (1170-1176 keV and 1330-1336 keV bands) and 
\Al\ the flux of the 1809 keV line
in the 1805-1813 keV band. The fluxes are obtained using the ``fitted-pattern'' method.}
\label{table:radiocativesbis}
\end{deluxetable}

\end{document}